\documentclass[pre,amsmath,amssymb,twocolumn]{revtex4}% Physical Review E
\usepackage{graphicx}% Include figure files
\usepackage{dcolumn}% Align table columns on decimal point
\usepackage{bm}% bold math
\usepackage{subfigure}
\usepackage{epsfig}
\usepackage{color}

\newcommand{\comment}[1]{}
\begin{document}

\title{Fixed Point Properties of the Ising Ferromagnet on the Hanoi Networks}
\author{S. Boettcher} 
\homepage{http://www.physics.emory.edu/faculty/boettcher/}
\author{C. T. Brunson}
\affiliation{Physics Dept., Emory University, Atlanta, GA 30322; USA}
\begin{abstract}
The Ising model with ferromagnetic couplings on the Hanoi networks
is analyzed with an exact renormalization group. In particular, the
fixed-points are determined and the renormalization-group flow for
certain initial conditions is analyzed. Hanoi networks combine a one-dimensional
lattice structure with a hierarchy of long-range bonds to create
a mix of geometric and small-world properties. Generically, those small-world
bonds result in non-universal behavior, i.e.~fixed points and scaling
exponents that depend on temperature and the initial choice of coupling
strengths. It is shown that a diversity of different behaviors can
be observed with seemingly small changes in the structure of the networks.
Defining interpolating families of such networks, we find tunable
transitions between regimes with power-law and certain essential singularities
in the critical scaling of the correlation length. These are similar to the
so-called inverted Berezinskii-Kosterlitz-Thouless transition previously
observed only in scale-free or dense networks. 
\end{abstract}
\maketitle

\section{Introduction\label{sec:Introduction}}

The study of equilibrium statistical models, in particular the Ising
model, on complex networks have led to a number of novel phenomena,
many of which still defy proper classification
\citep{Boccaletti06,Dorogovtsev08}.  The reasons for that originate
with the diversity of network structures conceivable, where the most
basic features, such as their degree distribution or average range of
connections, are insufficient to characterize their asymptotic
properties. Instead, even within a given ensemble of networks, scaling
behavior may depend strongly on the details of the network structure
or the strength of the interactions. Despite of that high degree
of non-universality, in fact, quite different types of networks
exhibit qualitatively very similar phenomena. That circumstance likely
points toward the possibility of generalized classification scheme
for statistical models in a networked world.  We will discuss the
possibilities of such a classification elsewhere \citep{Boettcher09b}.

One such ubiquitous phenomenon is the so-called inverted
Berezinskii-Kosterlitz-Thouless transition (BKT)
\citep{Dorogovtsev08}, named after the distinctly weak infinite-order
BKT phase transition observed for the $O(2)$-vector model on a
two-dimensional lattice
\citep{Berezinskii70,kosterlitz:73,Plischke94}.  In the original BKT
transition, spin-waves curl into vortex structures, whose
pair-creation and unbinding dominates the low-temperature behavior of
the system. Renormalization group analysis demonstrates the existence
of a line of stable fixed points in the low-$T$ regime that is
parameterized by the temperature $T$, resulting in an equally
$T$-dependent scaling behavior and an exponential singularity for the
correlation length there. Analogously, for several Ising spin systems
on certain networks, a similar $T$-dependent line of absorbing fixed
points has been found, but \emph{inverted} in the sense that it occurs
only in the \emph{high}-temperature regime, for all $T>T_{c}$
\citep{Bauer05,Hinczewski06,Hinczewski07,Khajeh07,Berker09}.  Clearly, such a
behavior can not be related to spin-waves due to the discrete,
$O(1)$-symmetry of Ising spins, nor to any topological defects such as
vortices due to the long-range nature of bonds in any of these
networks.

Here, we use an exact real-space renormalization group (RG) to study
the ferromagnetic Ising model on the recently introduced set of Hanoi
networks \citep{SWPRL,SWN,SWlong}. Hanoi networks have been used to
study phenomena as diverse as diffusion \citep{SWN,SWlong},
synchronization \citep{SWPRL}, the exclusion process \citep{TASEP09},
percolation \citep{Boettcher09c}, quantum transport
\citep{Boettcher11b}, and the vertex cover problem
\citep{Boettcher11c}. These hierarchically constructed networks
possess a regular degree or have an exponential degree distribution,
akin to the Watts-Strogatz Small Worlds \citep{Watts98}. Unlike
scale-free networks often used to model social phenomena
\citep{Barabasi99,Barabasi03}, they have a more {}``physically''
desirable geometry \citep{barthelemy_spatial_2010} of a lattice
backbone mixed with a small-world hierarchy.  With regard to spin
models, certain Hanoi networks might combine finite transition
temperatures with susceptibilities that exist in the high-temperature
regime. They hold the potential for an interesting, analytically
tractable interpolation from that small-world behavior toward that of
a finite dimensional lattice \citep{Kotliar83,Katzgraber03} by
re-weighting long-range bonds. As some of those features are lacking in
other hierarchical networks
\citep{Berker79,Andrade05,Hinczewski06,Andrade09}, this will serve as
focus of future investigations. For these Hanoi
networks, we determine the location of stable and unstable fixed
points in their phase diagrams and find a range of interesting
RG-flows.  Our results demonstrate that a scale-free network or highly
heterogeneous coupling strengths is not a prerequisite for the BKT
transition. Moreover, we provide analytically tractable one-parameter
interpolations between a ferromagnetic transition governed by unstable
fixed points and BKT-like transitions. In general, we find that the
critical divergence of the correlation length with temperature becomes
increasingly singular -- from power-law over BKT-like to a full
essential singularity -- for increasing relative strength of
long-range couplings in the network.

In the following Section, we describe the Hanoi networks. The analysis
of the phase diagrams and the RG-flow for the Ising ferromagnet on
these networks is discussed in Sec.~\ref{sec:Ising-Model}. In Sec.
\ref{sec:HN5y}, we introduce families of interpolating
networks to reveal a more comprehensive set of regimes, each with its
own characteristic type of phase transition, and we conclude with a
discussion of our results in their implications in
Sec.~\ref{sec:Conclusion}.

\section{Geometry of the Hanoi Networks\label{sec:Graph-Structure}}

\begin{figure}
\includegraphics[bb=43bp 14bp 822bp 450bp,clip,scale=0.3]{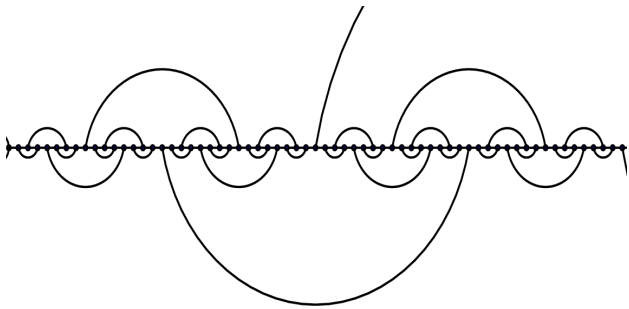}
\caption{(Color Online)\label{fig:3hanoi}
Depiction of the 3-regular network HN3 on a semi-infinite line.  }
\end{figure}

Each of the Hanoi networks possesses a simple geometric backbone, a
one-dimensional line of sites $n$, $0\leq n\leq N=2^{k}$
$(k\to\infty)$.  Each site is connected to its nearest neighbor,
ensuring the existence of the $1d$-backbone. To generate the
small-world hierarchy in these networks, consider parameterizing any
integer $n$ (except for zero) \emph{uniquely} in terms of two other
integers $(i,j)$, $i\geq0$, via
\begin{eqnarray}
n & = & 2^{i}\left(2j+1\right).
\label{eq:numbering}
\end{eqnarray}
Here, $i$ denotes the level in the hierarchy whereas $j\geq0$ labels
consecutive sites within each hierarchy. For instance, $i=0$ refers
to all odd integers, $i=1$ to all integers once divisible by 2 (i.e., 2, 6, 10,...), 
and so on. Depending on its level of the hierarchy,
any site has also small-world (i.e. long-range) bonds to more-distant
sites along the backbone, according to some non-random rule. For example,
we obtain a 3-regular network HN3 by connecting also 1 to 3, 5 to
7, 9 to 11, etc, for $i=0$, next 2 to 6, 10 to 14, etc, for $i=1$,
and 4 to 12, 20 to 28, etc., for $i=2$, and so on, as depicted in
Fig.~\ref{fig:3hanoi}. 

Previously\citep{SWPRL}, it was found that the average chemical path
between sites on HN3 scales as 
\begin{equation}
d^{HN3}\sim\sqrt{l}
\label{eq:3dia}
\end{equation}
with the distance $l$ along the backbone. In some ways, this property
is reminiscent of a square-lattice consisting of $N$ lattice sites.
The diameter (=diagonal) of this square is also $\sim\sqrt{N}$. 

\begin{figure}
\includegraphics[bb=100bp 100bp 400bp 700bp,clip,angle=-90,scale=0.4]{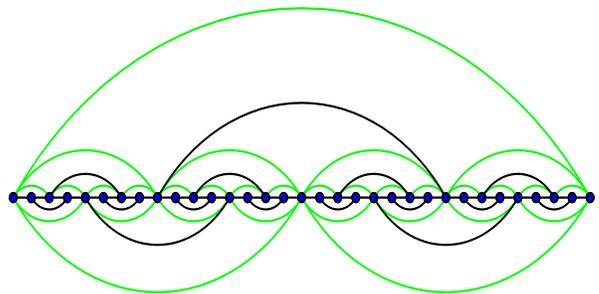}
\caption{(Color Online)\label{fig:5hanoi}Depiction of the planar network HN5, comprised
of HN3 (black lines) with the addition of further long-range bonds
(green-shaded lines). Note that sites on the lowest level of the hierarchy
have degree 3, then degree 5, 7, etc, comprising a fraction of $1/2,$
$1/4$, $1/8$, etc., of all sites, which makes for an average degree
5 in this network. }

\end{figure}

While HN3 (and HN4 \citep{SWPRL}) are of a fixed, finite degree,
we introduced here convenient generalizations of HN3 that lead to
new, revealing insights into small-world phenomena. First, we can
extend HN3 in the following manner to obtain a new planar network
of average degree 5, hence called HN5: In addition to the bonds in
HN3, in HN5 we also connect all even sites to both nearest sites \emph{within}
the same level of the hierarchy $i(\geq1)$. The resulting network
remains planar but now sites have a hierarchy-dependent degree, as
shown in Fig.~\ref{fig:5hanoi}. To obtain the average degree, we
observe that 1/2 of all sites have degree 3, 1/4 has degree 5, 1/8
has degree 7, and so on, leading to an exponentially falling degree
distribution of ${\cal P}\left\{ \alpha=2i+3\right\} \propto2^{-i}$.
Then, the total number of bonds $L$ in the system of size $N=2^{k}$
is
\begin{eqnarray}
2L & = & 2\left(2k-1\right)+\sum_{i=0}^{k-2}\left(2i+3\right)2^{k-1-i}\nonumber\\
 & = & 5\times2^{k}-8,
\label{eq:TotalLinksHN5}
 \end{eqnarray}
and thus, the average degree is 
\begin{eqnarray}
\left\langle \alpha\right\rangle  & = & \frac{2L}{N}\sim5.
\label{eq:averageDegreeHN5}
\end{eqnarray}

In HN5, the end-to-end distance is trivially 1, see Fig.~\ref{fig:5hanoi}.
Therefore, we define as the diameter the largest of the shortest paths
possible between any two sites, which are typically odd-index sites
furthest away from long-distance bonds. For the $N=32$ site network
depicted in Fig.~\ref{fig:5hanoi}, for instance, that diameter is
5, measured between site 3 and 19 (starting with $n=0$ as the left-most
site), although there are many other such pairs. It is easy to show
recursively that this diameter grows as
\begin{eqnarray}
d^{HN5} & = & 2\left\lfloor \frac{k}{2}\right\rfloor +1\sim\log_{2}N.
\label{eq:5dia}
\end{eqnarray}
We have checked numerically that the \emph{average} shortest path
between any two sites appears to increase logarithmically with system
size $N$ as well.

\begin{figure}
\includegraphics[bb=0bp 60bp 760bp 450bp,clip,scale=0.3]{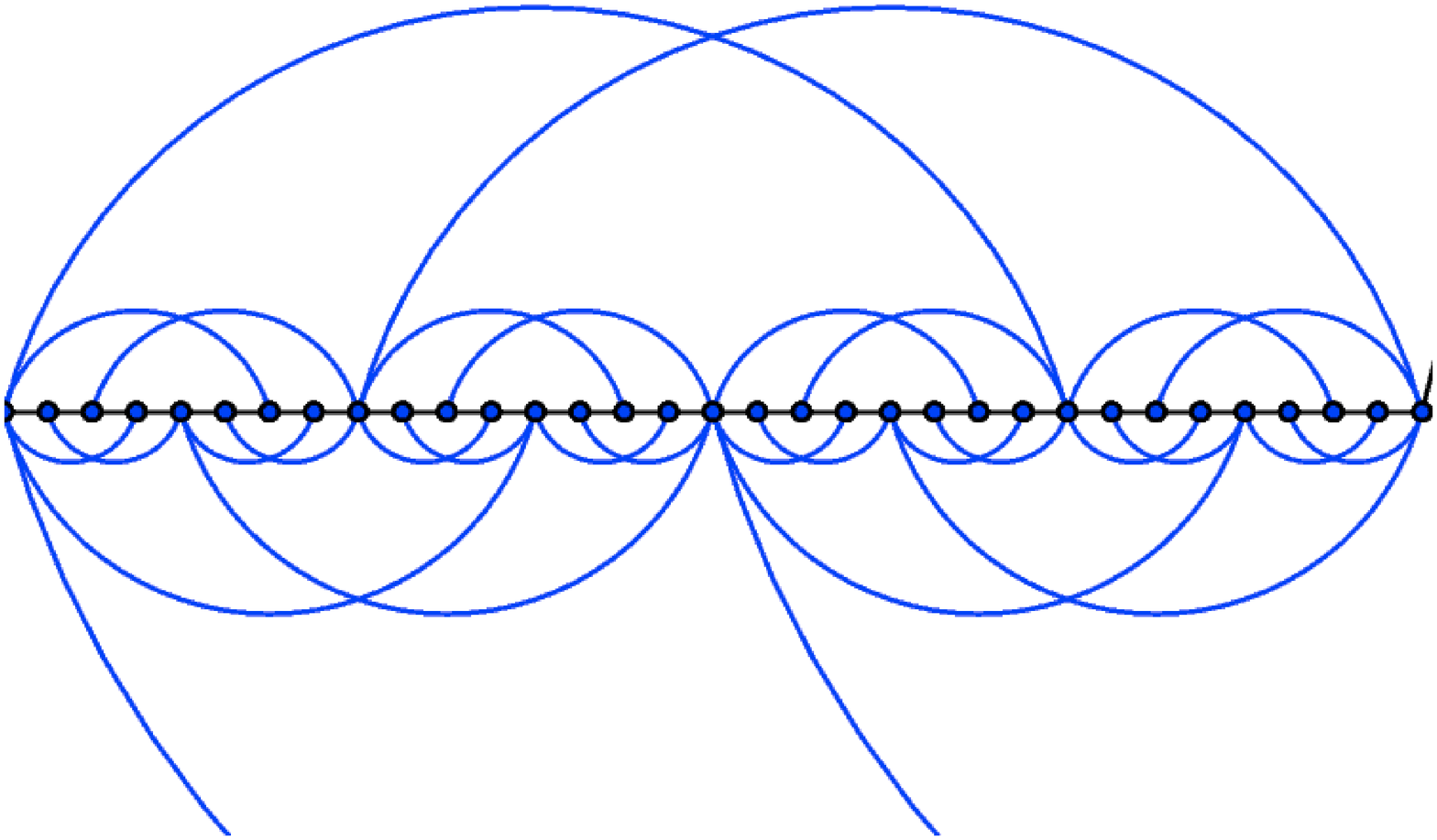}
\caption{(Color Online)\label{fig:hanoi_nonplanar}Depiction of the non-planar Hanoi network
HNNP. Again, starting from a $1d$-backbone (black lines), a set of
long-range bonds (blue-shaded lines) is added that break planarity
but maintain the hierarchical pattern set out in Eq.~(\ref{eq:numbering}).
The RG on this network remains exact. Note that sites on the lowest
two levels of the hierarchy have degree 3, then degree 5, 7, etc,
comprising a fraction of $1/2,$ $1/4$, $1/8$, etc., of all sites,
which makes for an average degree of 4 in this network. }
\end{figure}

The networks HN3 and HN5 have the convenient but (from a mean-field
perspective) unrealistic restriction of being planar. In fact, with
a minor extension of the definition, it is easy to also design Hanoi
networks that are both, non-planar \emph{and} fully renormalizable.
The simplest such network, which we dub HNNP, is depicted in Fig.~\ref{fig:hanoi_nonplanar}. 

To obtain the average degree, we observe that $1/2+1/4$ of all sites
have degree 3, $1/8$ has degree 5, $1/16$ has degree 7, and so on,
leading to an exponentially falling degree distribution, as for HN5.
The total number of bonds $L$ in the system of size $N=2^{k}$ is
\begin{eqnarray}
2L & = & 3\times2^{k-1}+\sum_{i=2}^{k-1}\left(2i-1\right)2^{k-i}+2k+(2k-2),\nonumber\\
 & = & 4\times2^{k}-4,
\label{eq:TotalLinksHN-NP}
\end{eqnarray}
and thus, the average degree is 
\begin{equation}
\left\langle \alpha\right\rangle   =  \frac{2L}{N}\sim4.
\label{eq:averageDegreeHN-NP}
\end{equation}
Here, too, it is easy to see that the shortest paths between sites
increases logarithmically with system size $N$. Note that in the
same manner we extended HN3 to HN5 by connecting even-indexed sites
within each hierarchy, we will discuss such an extension of HNNP toward
a non-planar network of average degree 6, HN6, in Sec.~\ref{sub:Analysis-of-HN6:}.

\section{RG for the Ising Model\label{sec:Ising-Model}}

In this section, we study Ising spin models on HN3, HN5 and HNNP with
the renormalization group (RG) \citep{Plischke94}. First, we consider
RG for the Ising model on HN3. This allows us to introduce our procedure
although the result turns out to be trivial in the sense that there
is no finite-temperature transition. Especially, it is actually identical
(except for one bond-type) to the exact RG for HN5, and it is almost
identical to the treatment below for HNNP.

\begin{figure}
\includegraphics[bb=0bp 550bp 400bp 750bp,clip,scale=0.5]{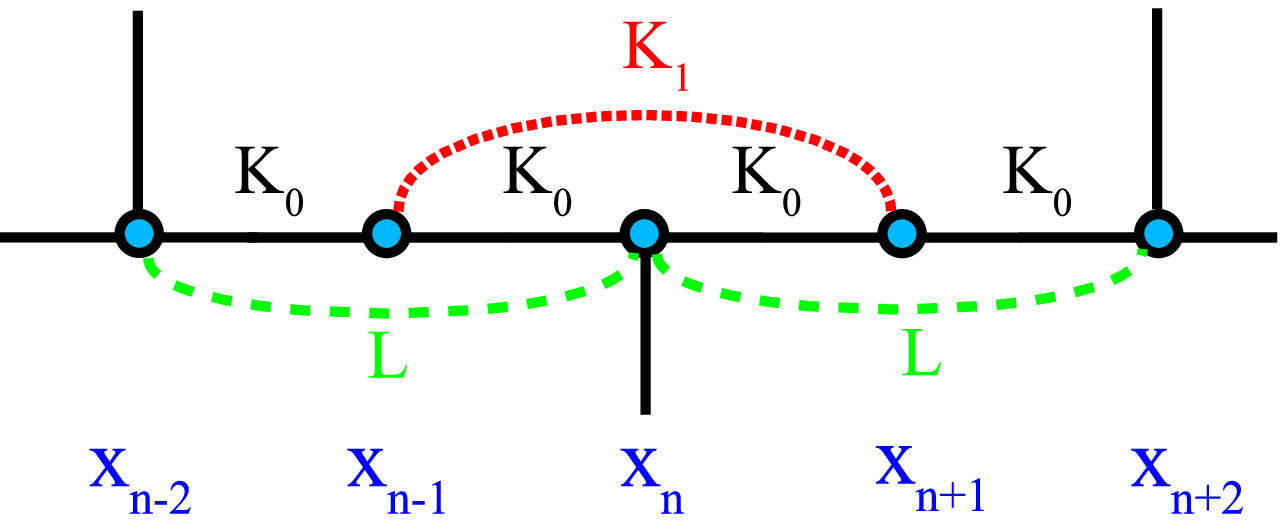}
\includegraphics[bb=50bp 550bp 320bp 750bp,clip,scale=0.5]{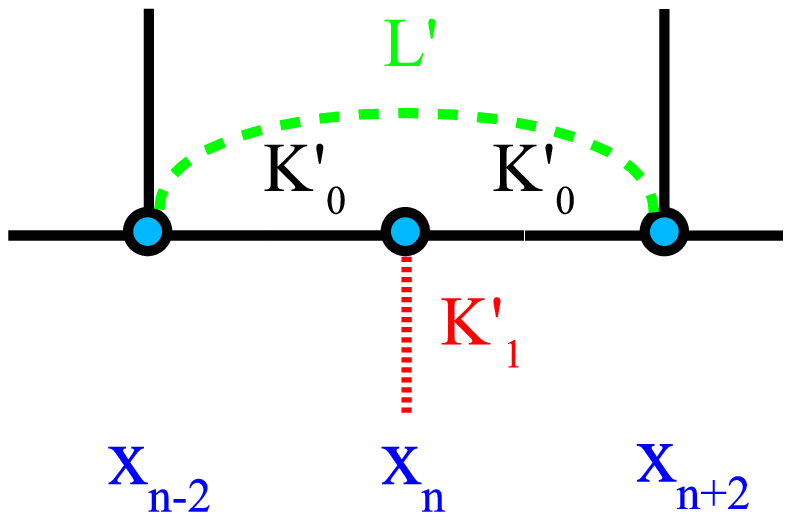}
\caption{(Color Online)\label{fig:RG3}Depiction of the (exact) RG step for the Ising model
on HN3. The step consists of tracing out odd-labeled variables $x_{n\pm1}$
in the top plot and expressing the renormalized couplings $(L'_0,K'_{0})$
on the bottom in terms of the old couplings $(L_0,K_{0},K_{1})$. Note
that the original network in Fig.~\ref{fig:3hanoi} does not contain
couplings of type $(L_0,L'_0)$, but that they certainly become relevant
during the process. }
\end{figure}

\subsection{Ising Ferromagnet on HN3\label{sub:Ising-Ferromagnet-on-HN3}}

The RG consists of recursively tracing out spins level-by-level in
the hierarchy \citep{SWPRL}. In terms of Eq.~(\ref{eq:numbering}),
we start by tracing out all sites with $n$ odd, i.e. $i=0$, then
those $n$ which are divisible by 2 only once, i.e. $i=1$, and so
on. We can always relabel all sites $n$ after any RG step by $n\to n/2$,
so that we trace out the respective odd-relabeled sites at any level.
It is apparent, for instance from Fig.~\ref{fig:3hanoi}, that odd-labeled
sites are connected to their even-labeled nearest neighbors on the
backbone, say, by a coupling $K_{0}\left(=\beta J_{0}\right)$. At
any level, each odd-labeled site $x_{n\pm1}$ is also connected to
one other such site $x_{n\mp1}$ across an even-labeled site $x_{n}$
with $n=2\left(2j+1\right)$ that is exactly \emph{once} divisible
by 2. Let us call that coupling $K_{1}\left(=\beta J_{1}\right)$.
The basic RG step is depicted in Fig.~\ref{fig:RG3} and consists
of tracing out the two sites $x_{n\pm1}$ neighboring the site $x_{n}$
for all $j$ with $n=2\left(2j+1\right)$. 

We can section the Ising Hamiltonian
\begin{eqnarray}
-\beta\mathcal{H} & = &\sum_{n=1}^{2^{k-2}}\left(-\beta\mathcal{H}_{n}\right)
+\mathcal{R}\left(K_{2},K_{3},\ldots\right),
\label{eq:3Hamiltonian}
\end{eqnarray}
where $\mathcal{R}$ contains all coupling terms of higher level in
the hierarchy, and each sectional Hamiltonian is given by
\begin{eqnarray}
&&-\beta\mathcal{H}_{n}=4I+L_{0}\left(x_{n-2}x_{n}+x_{n}x_{n+2}\right)+ \nonumber\\
 & & K_{0}\left(x_{n-2}x_{n-1}+x_{n-1}x_{n}+x_{n}x_{n+1}+x_{n+1}x_{n+2}\right)\nonumber \\
 &  &\qquad +K_{1}x_{n-1}x_{n+1},
\label{eq:3HSection}
\end{eqnarray}
where $\left(K_{0},K_{1},L_0\right)$ are the unrenormalized couplings
defined in Fig.~\ref{fig:RG3} and $I$ is a constant that fixes the
overall energy scale per spin. (There are effectively 4 spins involved
in each graph-let, as those at each boundary are equally shared with
neighboring graph-lets.) While couplings of the type $L_{0}$ between
next-nearest even-labeled neighbors emerges that are not part of the
network initially in HN3, they do emerge during the RG step (otherwise
the system of recursion equations would not close), see Fig.~\ref{fig:RG3}. 

To simplify the analysis, we introduce new variables similar to inverse
{}``activities'' \citep{Plischke94},
\begin{equation}
C=e^{-4I},\,\,\kappa=e^{-4K_{0}},\,\,\lambda=e^{-4L_{0}},\,\,\mu=e^{-2K_{1}},
\label{eq:activities}
\end{equation}
which ensure that the RG flow only contains algebraic functions and,
for the ferromagnetic model, remains confined within the physical
domain $0\leq\kappa,\lambda,\mu\leq1$. Thus, we rewrite Eq.~(\ref{eq:3HSection})
as
\begin{eqnarray}
e^{-\beta\mathcal{H}_{n}} & = & C^{-1}\kappa^{-\frac{1}{4}\left(x_{n-2}x_{n-1}+x_{n-1}x_{n}+x_{n}x_{n+1}+x_{n+1}x_{n+2}\right)}\nonumber \\
 &  &\quad\lambda^{-\frac{1}{4}\left(x_{n-2}x_{n}+x_{n}x_{n+2}\right)}\mu^{-\frac{1}{2}x_{n-1}x_{n+1}}.
\label{eq:3HSectPower}
\end{eqnarray}
 
Tracing out the odd-labeled spins, we have to evaluate 
\begin{eqnarray}
 &  & \sum_{\left\{ x_{n-1}=\pm1\right\} }\sum_{\left\{ x_{n+1}=\pm1\right\} }e^{-\beta\mathcal{H}_{n}}\nonumber \\
 & = & C^{-1}\mu^{-\frac{1}{2}}\lambda^{-\frac{1}{4}\left(x_{n-2}x_{n}+x_{n}x_{n+2}\right)}\label{eq:3trace-1}\\
 &  & \quad\left[\kappa^{-\frac{1}{4}\left(x_{n-2}+2x_{n}+x_{n+2}\right)}+\mu\kappa^{-\frac{1}{4}\left(x_{n-2}-x_{n+2}\right)}\right.\nonumber \\
 &  & \qquad\left.+\mu\kappa^{\frac{1}{4}\left(x_{n-2}-x_{n+2}\right)}+\kappa^{\frac{1}{4}\left(x_{n-2}+2x_{n}+x_{n+2}\right)}\right]\nonumber \\
 & = & \left(C'\right)^{-\frac{1}{2}}\left(\lambda'\right){}^{-\frac{1}{4}x_{n-2}x_{n+2}}\left(\kappa'\right){}^{-\frac{1}{4}\left(x_{n-2}x_{n}+x_{n}x_{n+2}\right)}\nonumber 
\end{eqnarray}
for the remaining spins in terms of the \emph{renormalized} quantities
$C',\kappa',\lambda'$. Of the eight possible relations resulting
from the combinations $x_{n-2},x_{n},x_{n+2}=\pm1$, only three are
independent. After some algebra, we extract from those the RG recursions:
\begin{eqnarray}
\kappa' & = & \kappa\lambda\frac{2\left(1+\mu\right)}{1+2\mu\kappa+\kappa^{2}},\nonumber \\
\lambda' & = & \frac{\left(1+\kappa\right)^{2}\left(1+\mu\right)}{2\left(1+2\mu\kappa+\kappa^{2}\right)},\label{eq:3RG}\\
C' & = & C^{2}\frac{\kappa\mu}{\sqrt{2}\left(1+\kappa\right)\left(1+\mu\right)^{\frac{3}{2}}\sqrt{1+2\mu\kappa+\kappa^{2}}}.\nonumber 
\end{eqnarray}
Note that for couplings in higher levels of the hierarchy it is $K'_{i}=K_{i+1}$
for $i\geq1$; correspondingly, these couplings, and hence, $\mu$,
will \emph{not} renormalize. Instead, they retain their {}``bare''
value $\mu^{2}$ determined by the temperature, $kT/J=-2/\ln\mu$.
In this sense, we will use $\mu$ as a measure of temperature throughout. 

\begin{figure}
\includegraphics[clip,scale=0.33]{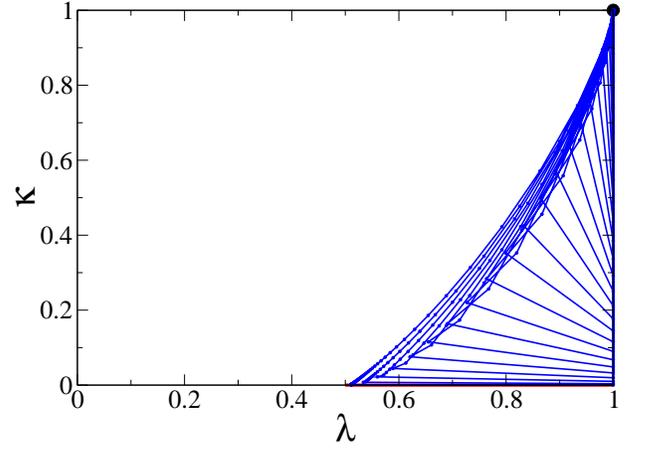}
\caption{(Color Online)\label{fig:IsingPhaseHN3}Phase diagram for the Ising ferromagnet
on HN3. The stable high-temperature fixed point at $\kappa^{*}=\lambda^{*}=1$
is marked by a black dot, the line of unstable fixed points at low
temperature by a thick red line for $\kappa^{*}=0$ and $\frac{1}{2}\leq\lambda^{*}\leq1$.
For homogeneous initial conditions in Eq.~(\ref{eq:HN3activitiesIC}),
the ensuing RG flow for each $\mu$ (blue lines) starts on the line
$\lambda^{(0)}=1$ and evolves to a higher value of $\kappa^{(1)}$
but smaller value of $\lambda^{(1)}$ before it rapidly veers off
toward the high-temperature fixed point. RG flow lines are drawn
here for equal increments in $\mu=\sqrt{\kappa^{(0)}}$ . Note that
lines are allowed to cross, as each flow line is parameterized by
a different value of $\mu$.}
\end{figure}

Only half of the contribution to the renormalized energy scale is
originating with the sectional Hamiltonian in
Eq.~(\ref{eq:3HSection}), since at the next level \emph{two} such
sections are combined into one, making $C'\propto C^{2}$. While we do
not consider the recursions for $C$ in this paper, they are essential
to reconstruct the free energy for each system, and will be analyzed
elsewhere \citep{Boettcher09b}.

Eqs.~(\ref{eq:3RG}) provide recursions order-by-order in the RG for
the evolution of the effective couplings characterizing increasingly
larger scales of the network. To facilitate this RG flow, we need to
specify initial conditions for a particular physical situation
realized in the unrenormalized, bare network. Here, we restrict
ourselves to networks with uniform bonds (although many interesting
choices are conceivable, such as distance-dependence
\citep{Kotliar83,Katzgraber03,Hinczewski06}).  For HN3 this implies
that we chose $J=1$ as our energy scale, such that $K_{i}=\beta
J=\beta$ and $I=L_{0}=0$ initially, or in terms of
Eq.~(\ref{eq:activities}),
\begin{equation}
C^{(0)}=\lambda^{(0)}=1,\qquad\kappa^{(0)}=\mu^{2}=e^{-4\beta}.
\label{eq:HN3activitiesIC}
\end{equation}

Searching for fixed points $K'_{0}=K_{0}=K_{0}^{*}$ and $L'=L=L^{*}$,
i.e. $\kappa'=\kappa=\kappa^{*}$ and $\lambda'=\lambda=\lambda^{*}$
in Eqs.~(\ref{eq:3RG}), immediately provides the trivial, high-temperature
solution $\kappa^{*}=\lambda^{*}=1$, i.e. $K_{0}^{*}=L^{*}=0$.
Further analysis yields only a line of (unstable) strong-coupling
fixed points, 
\begin{equation}
\kappa^{*}=0,\quad\lambda^{*}=\frac{1+\mu}{2},
\label{eq:3RGfixedpoint}
\end{equation}
extending from $\lambda^{*}=\frac{1}{2}$ for low temperatures, $\mu=0$,
to $\lambda^{*}=1$ for $T\to\infty$, where $\mu=1$, see Fig.~\ref{fig:IsingPhaseHN3}.
Even at $T=0$, only the renormalized backbone bonds $K_{0}$ provide
strong coupling, the emerging long-range bonds $L_{0}$ only exert
limited coupling strength.

Local analysis near the fixed points with the Ansatz
\begin{eqnarray}
\kappa_{n} & \sim &
\kappa^{*}+\epsilon_{n},\quad\lambda_{n}\sim\lambda^{*}+\delta_{n},\quad\epsilon_{n},\delta_{n}\ll1
\label{eq:Ansatzdeleps}
\end{eqnarray}
reveals that the high-temperature fixed point is always stable and
corrections decay exponentially, where the exponential contains a
factor of $2^{\frac{n}{2}}=\sqrt{N}$. At the low-temperature line
of fixed points in Eq.~(\ref{eq:3RGfixedpoint}) we find 
\begin{eqnarray}
\epsilon_{n}\sim\epsilon_{0}\left(1+\mu\right)^{2n}, &  &
\delta_{n}\sim\frac{1-\mu}{1+\mu}\epsilon_{0}\left(1+\mu\right)^{2n},
\label{eq:HN3correction}
\end{eqnarray}
which is divergent for all $T>0$, i.e. $0<\mu\leq1$, making the
fixed point at $T=0$ unstable. For any fixed point, there is no linear
expansion possible that would yield critical exponents. For the initial
conditions in Eqs.~(\ref{eq:HN3activitiesIC}) corresponding to uniform
couplings throughout the unrenormalized network, the RG flow always
evolves to the high-temperature fixed point, as Fig.~\ref{fig:IsingPhaseHN3}
shows. Thus, the ferromagnet on this network behaves similar to a
$1d$ Ising model.

\begin{figure}
\includegraphics[bb=0bp 550bp 400bp 750bp,clip,scale=0.5]{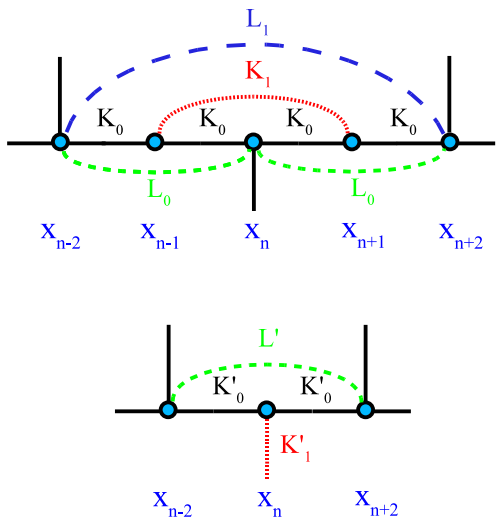}
\includegraphics[bb=50bp 550bp 330bp 750bp,clip,scale=0.5]{RG3hanoi_after}
\caption{(Color Online)\label{fig:RG5}Depiction of the (exact) RG step for the Ising model
on HN5. This step is $identical$ to that for HN3 in Fig.~\ref{fig:RG3}
aside from the extra link $L_{1}$ spanning between $x_{n-2}$ and
$x_{n+2}$ (top), which contributes to the renormalization of $L'_{0}$
(bottom).}
\end{figure}

\subsection{Ising Ferromagnet on HN5\label{sub:Ising-Ferromagnet-on-HN5}}

As shown in Sec.~\ref{sec:Graph-Structure}, HN5 is basically an extension
of HN3, created by adding a new layer of links to each level of the
hierarchy. As is apparent from the foregoing discussion in Sec.~\ref{sub:Ising-Ferromagnet-on-HN3},
these additions correspond precisely to new renormalizable operators
(here, the bonds $L$) that \emph{inevitably} \emph{emerge} during
the RG of HN3, see Fig.~\ref{fig:RG3}. In HN5, these new operators
are simply deemed an original feature of the network, hence, maintaining
the RG as an exact procedure. Consequently, the RG itself hardly changes,
see Fig.~\ref{fig:RG5}; it merely differs by one extra link in the
graph-let, $L_{1}$, compared to that for HN3 in Fig.~\ref{fig:RG3}.
In Eq.~(\ref{eq:3HSection}), it only adds the term $L_{1}x_{n-2}x_{n+2}$
to the sectional Hamiltonian and, like $L_{0}$ itself, $L_{1}$ does
not get traced over in the calculation in Eq.~(\ref{eq:3trace-1}).
We can introduce these new bonds as yet another free, non-renormalizing
coupling in the RG and choose, to wit, 
\begin{eqnarray}
L_{1}=yK_{1}, & \quad{\rm i.e.\quad} & e^{-4L_{1}}=\mu^{2y}.
\label{eq:L1bond}
\end{eqnarray}
This merely contributes a factor of $\mu{}^{-\frac{y}{4}x_{n-2}x_{n+2}}$
to the unprimed side of Eq.~(\ref{eq:3trace-1}), which correspondingly
alters only the recursion for $\lambda'$ in Eq.~(\ref{eq:3RG}) by
a factor of $\mu^{2y}$. Otherwise using the same definitions as in
Sec.~\ref{sub:Ising-Ferromagnet-on-HN3}, we obtain the RG recursions
for the Ising ferromagnet on HN5:%
\footnote{Absent the $K_{i}$ bonds for $i>0$, these relations trivially reproduce
the $1d$-hierarchical lattice with small-world bonds.%
}
\begin{eqnarray}
\kappa' & = & \kappa\lambda\frac{2\left(1+\mu\right)}{1+2\mu\kappa+\kappa^{2}},\nonumber \\
\lambda' & = & \mu^{2y}\,\frac{\left(1+\kappa\right)^{2}\left(1+\mu\right)}{2\left(1+2\mu\kappa+\kappa^{2}\right)},\label{eq:5RG}\\
C' & = & C^{2}\frac{\kappa\mu}{\sqrt{2}\left(1+\kappa\right)\left(1+\mu\right)^{\frac{3}{2}}\sqrt{1+2\mu\kappa+\kappa^{2}}}.\nonumber 
\end{eqnarray}
Accordingly, due to the bare existence of the $L_{1}$ bond, we will
have to change the initial conditions from
Eq.~(\ref{eq:HN3activitiesIC}) to
\begin{eqnarray}
C^{(0)} & = &1,\quad\kappa^{(0)}=\mu^{2}=e^{-4\beta},\quad\lambda^{(0)}=\mu^{2y}.
\label{eq:HN5activitiesIC}
\end{eqnarray}

Analyzing these recursions for fixed points, $\kappa'=\kappa=\kappa^{*}$
and $\lambda'=\lambda=\lambda^{*}$, we find that the addition of
the extra long-range bond has \emph{eliminated} the high-temperature
fixed point found in HN3. At low temperatures, we find similar to
Eq.~(\ref{eq:3RGfixedpoint}) in HN3 a line of fixed points 
\begin{equation}
\kappa^{*}=0,\quad\lambda^{*}=\frac{\mu^{2y}}{2}\left(1+\mu\right),
\label{eq:5RGfixedpointT0}
\end{equation}
which here extends over the entire domain for the long-range bonds,
$0\leq\lambda^{*}\leq1$ for $0\leq\mu\leq1$. Note that although
$y$ represents a continuous interpolation between HN3 and HN5, there
is a singular limit at $y\to0$ toward an isolated point corresponding
to HN3, see Eqs.~(\ref{eq:3RGfixedpoint}). In the following, we only
treat the case of couplings that are homogeneous throughout the unrenormalized
network, $y=1$. Consideration of the rich set of transitions occurring
for the family of networks parameterized by interpolating $0<y\leq1$
is deferred to Sec.~\ref{sec:HN5y}.

\begin{figure}
\includegraphics[clip,scale=0.33]{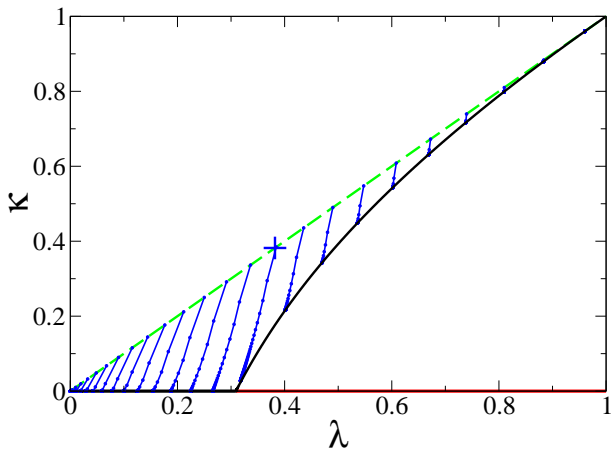}
\caption{(Color Online)\label{fig:IsingPhaseHN5}Phase diagram for the Ising ferromagnet
on HN5 with homogeneous bonds, $y=1$. Unlike for HN3 in Fig.~\ref{fig:IsingPhaseHN3},
there is no high-temperature fixed point here, instead a line of stable
fixed points marked by a thick black line that extends from the strong-coupling
regime at $\kappa^{*}=0$ and $0\leq\lambda^{*}<\frac{1}{2\phi}=0.309017\ldots$
all the way to the high-temperature regime. For the homogeneous-coupling
initial conditions in Eq.~(\ref{eq:HN5activitiesIC}), the ensuing
RG flow (blue-shaded lines) starts on the diagonal (dashed line) and
always evolves toward lower values of $\kappa^{*}$ and $\lambda^{*}$.
In the low-temperature regime, the flow reaches strong coupling, $\kappa^{*}=0$,
for values below $\mu_{c}=\phi^{-1}=0.618033\ldots$ (cross at $\kappa^{(0)}=\lambda^{(0)}=\mu_{c}^{2}$).
Above that, the flow terminates on the line of finite-coupling fixed
points, all the way to infinite temperature, $\mu\to1$.}
\end{figure}

Dividing out the $\kappa^{*}=0$-solution, further analysis of Eqs.
(\ref{eq:5RG}) for $y=1$ reveals yet another line of fixed points
given by 
\begin{eqnarray}
\kappa^{*} & = & \frac{1}{2}\left[-\left(1-\mu\right)\mu+\sqrt{\left(1+\mu\right)\left(\mu^{3}-3\mu^{2}+8\mu-4\right)}\right],\nonumber \\
\label{eq:5RGfpl}\\
\lambda^{*} & = & \frac{\mu}{4}\left[2-\mu+\mu^{2}+\sqrt{\left(1+\mu\right)\left(\mu^{3}-3\mu^{2}+8\mu-4\right)}\right],\nonumber 
\end{eqnarray}
which can be expressed most simply in closed form as
\begin{equation}
\lambda^{*}=\frac{1}{2}\left[\kappa^{*}-1+\sqrt{5+2\kappa^{*}+5\left(\kappa^{*}\right){}^{2}+4\left(\kappa^{*}\right){}^{3}}\right]
\label{eq:lamkap_fpl}
\end{equation}
by eliminating $\mu$. As we will see, these relations lead to physical
fixed points only within a limited range of the temperature $\mu$.
We have plotted the phase diagram for HN5 at $y=1$ in Fig.~\ref{fig:IsingPhaseHN5}.
It also shows the RG flow for the initial conditions in Eqs.~(\ref{eq:HN5activitiesIC}),
which starts on the diagonal, representing all-equal bonds for the
homogeneous network. For these initial conditions, the flow always
evolves toward smaller values of $\kappa$ and $\lambda$, i.e.
stronger coupling. But there is a notable transition where the attained
fixed point jumps from the low-temperature branch in Eq.~(\ref{eq:5RGfixedpointT0})
characterized by $\kappa^{*}=0$, i.e. a solidly frozen backbone,
to the branch given Eq.~(\ref{eq:lamkap_fpl}) on which both $\kappa^{*}$
and $\lambda^{*}$ are finite. We obtain this transition point by
evaluating Eqs.~(\ref{eq:5RGfpl}-\ref{eq:lamkap_fpl}) for $\kappa^{*}=0$,
which yields $\lambda^{*}=\frac{1}{2\phi}=0.309017\ldots$ and
a critical temperature of
\begin{equation}
\mu_{c}=\frac{1}{\phi}, \enskip{\rm or}\enskip 
\frac{kT_{c}}{J}=-\frac{2}{\ln\mu_{c}}=4.156173842\ldots,
\label{eq:HN5Tc}
\end{equation}
where $\phi=\left(\sqrt{5}+1\right)/2=1.6180339887\ldots$ is the
{}``golden ratio'' \citep{SWN}. For bare couplings at this temperature,
marked by a cross in Fig.~\ref{fig:IsingPhaseHN5}, the RG flow marginally
reaches the strong-coupling limit. In this network, for these initial
conditions the RG flow never reaches an unstable fixed point such
as the unstable portion of Eq.~(\ref{eq:5RGfixedpointT0}), marked
by a red-shaded line in Fig.~\ref{fig:IsingPhaseHN5}. As we will
see below, this circumstance will change when we weaken the impact
of long-range couplings.

\begin{figure}
\includegraphics[clip,scale=0.33]{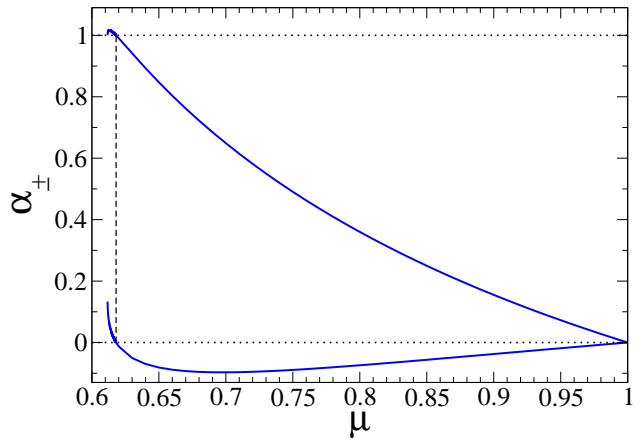}
\caption{(Color Online)\label{fig:IsingHN5ev}Plot of the two eigenvalues $\alpha_{\pm}$
(blue-shaded lines) resulting from Eq.~(\ref{eq:HN5matrix}). In
the physical regime, $\frac{1}{\phi}<\mu\leq1$, marked left by the
vertical dashed line, both eigenvalues satisfy $1>\alpha_{+}\geq0\geq\alpha_{-}>-1$.
Just below that regime, $\alpha_{+}$ exceeds unity, re-enters, and
both eigenvalues soon (at $\mu=0.6117\ldots$) separately disappear
into the complex plane.} 
\end{figure}

\subsubsection{Fixed-Point Stability \label{sub:Fixed-Point-Stability-HN5} }

We determine the stability of the fixed points with a local
analysis using  Eq.~(\ref{eq:Ansatzdeleps}). At $y=1$,
inserted into the strong-coupling solution, Eq.~(\ref{eq:5RGfixedpointT0}),
provides
\begin{eqnarray}
\epsilon_{n+1} & = & \mu^{2}\left(1+\mu\right)^{2}\epsilon_{n}
\label{eq:epsHN5}
\end{eqnarray}
and $\delta_{n}\propto\epsilon_{n}$. Thus, this line of fixed points
is stable only for $\mu$ smaller than the solution of $1=\mu\left(1+\mu\right)$,
i.e. $0\leq\mu<\phi^{-1}$; for larger values of $\mu$ the line
is unstable. In terms of $\lambda^{*}$, this implies that the stability
of the line of strong coupling fixed points changes exactly at
the location in the phase diagram, Fig.~\ref{fig:IsingPhaseHN5}, where
the line of high-temperature fixed points in Eq.~(\ref{eq:lamkap_fpl})
intersects, at $\left(\frac{1}{2\phi},0\right)$. Applying the Ansatz
on the line of high-temperature fixed points in Eqs.~(\ref{eq:5RGfpl})
yields after some algebra 
\begin{eqnarray}
\epsilon_{n+1} & = & -\frac{1}{2\mu}\left(2-\mu-\sqrt{\frac{\mu^{3}-3\mu^{2}+8\mu-4}{1+\mu}}\right)\,\epsilon_{n}\nonumber \\
 & + & \frac{2-3\mu-\mu^{2}+\sqrt{\left(1+\mu\right)\left(\mu^{3}-3\mu^{2}+8\mu-4\right)}}{2\left(1-\mu\right)\mu}\,\delta_{n},\nonumber \\
\delta_{n+1} & = &\frac{\mu}{2}-\frac{1}{2}\sqrt{\frac{\mu^{3}-3\mu^{2}+8\mu-4}{1+\mu}}\,\epsilon_{n}.
\label{eq:HN5matrix}
\end{eqnarray}
The two eigenvalues $\alpha_{\pm}$ characterizing this system are
easy to find but lead to messy algebraic expressions; instead of presenting
them, we merely plot their values as a function of $\mu$ in Fig.
\ref{fig:IsingHN5ev}. It shows that instability and a delicate root-singularity
(at $\mu=1+\frac{1}{3}\sqrt[3]{6\sqrt{114}-27}-5/\sqrt[3]{6\sqrt{114}-27}=0.6117\ldots$)
lurk just outside the physical regime, $\phi^{-1}<\mu\leq1$, yet
inside of that regime the eigenvalues both satisfy $\left|\alpha_{\pm}\right|<1$
and the entire line of fixed points remains stable.

\subsubsection{Divergence of the Correlation Length at $T_{c}$\label{sub:Exponential-Divergence-of}}

The effect of the critical point at $\mu_{c}=\frac{1}{\phi}$ on the
scaling behavior of the Ising model deserves special attention.
For instance, we can associate a correlation length $\xi$ at a given
temperature $\mu$ to the system by considering the approach to the
stable fixed point, which is derived from the dominant eigenvalue
$\alpha_{+}(\mu)<1$ of Eq.~(\ref{eq:HN5matrix}) affecting the asymptotic
behavior of the backbone couplings $\kappa_{n}$,\[
\epsilon_{n}\sim\left(\alpha_{+}\right)^{n}\epsilon_{0}\sim
e^{-\frac{n}{n^{*}}},\] defining a cut-off scale $n^*$.
Aside from a few initial transient rescalings, associated with local
structure of the network, the appearance of the physical state of
the system remains scale invariant, $\epsilon_{n+1}\sim\epsilon_{n}$,
for system sizes $N=2^{n}$ with $n<n^{*}$. In turn, for larger systems
with $n>n^{*}$, correlated domains of size 
\begin{equation}
\xi\left(\mu\right)\sim2^{n^{*}}=e^{\frac{\ln2}{-\ln\alpha_{+}}}
\label{eq:def_xi}
\end{equation}
become mutually decorrelated to attest to the off-critical macroscopic
state of the system. Only for $\mu\to\mu_{c}$ does the correlation
length diverge: Expanding the eigenvalue resulting from
Eqs.~(\ref{eq:HN5matrix}) for $\mu\to\mu_{c}^+$ , we obtain
$\alpha_{+}\sim1-a\left(\mu-\mu_{c}\right)$ with $a=2\sqrt{5}$. In
fact, the corresponding expansion around the strong-coupling fixed
point in Eq.~(\ref{eq:5RGfixedpointT0}) for $\mu\to\mu_{c}^-$ , we
obtain $\alpha_{+}\sim1-a\left(\mu_{c}-\mu\right)$ with the same
constant $a$, so that we can write
$\alpha_{+}\sim1-a\left|\mu-\mu_{c}\right|$ and get
\begin{equation}
\xi\sim e^{\frac{\ln2}{a\left|\mu-\mu_{c}\right|}},\qquad\mu\to\mu_{c}.
\label{eq:xi_exp-divergence}
\end{equation}
This exponential divergence neither resembles the power-law divergence
for a finite-order phase transition of Ising systems on a lattice
nor the infinite-order transition characteristic of BKT that we
discuss below.

\subsection{Ising Ferromagnet on HNNP\label{sub:Ising-Ferromagnet-on-HNNP}}

\begin{figure}
\includegraphics[bb=0bp 0bp 369bp  200bp,clip,scale=0.5]{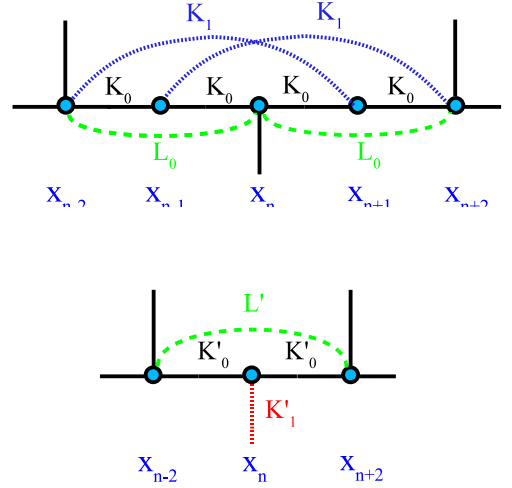}
\includegraphics[bb=50bp 550bp 330bp 750bp,clip,scale=0.5]{RG3hanoi_after}
\caption{(Color Online)\label{fig:RGHN-NP}Depiction of RG step for the Ising model on HNNP.
This step differs from that for HN3 in Fig.~\ref{fig:RG3} and HN5
in Fig.~\ref{fig:RG5} as there are a set of crossing long-range bonds
in each elementary graph-let before the RG step (left). But the resulting
graph-let after the RG step is again identical to that for HN3 and
HN5. }
\end{figure}

HNNP represents a drastic change in the geometric properties over the
other Hanoi networks, HN3 and HN5, while nonetheless preserving the
exact RG. Here, a crossing set of bonds $K_{1}$ in the elementary
graph-let shown in Fig.~\ref{fig:RGHN-NP} render this network
non-planar.  Yet, its average degree of 4 is between that for HN3 and
HN5, and we will see that the effective strength of the renormalized
coupling is intermediate between HN3 and HN5, too: Unlike HN3, it does
have an ordered low-temperature regime, but unlike for the transition
in HN5, its high-temperature regime above $T_{c}$ only possesses a
high-temperature fixed point of vanishing coupling strength; not
enough to sustain the partially ordered state found in HN5.

The structural feature of the crossing bonds is reflected in the Ising
Hamiltonian, of course, and we find for its sectional Hamiltonian
\begin{eqnarray}
&&-\beta\mathcal{H}_{n} =4I+L_{0}\left(x_{n-2}x_{n}+x_{n}x_{n+2}\right)+\nonumber \\
 &&K_{0}\left(x_{n-2}x_{n-1}+x_{n-1}x_{n}+x_{n}x_{n+1}+x_{n+1}x_{n+2}\right)\nonumber \\
 &  & \quad+K_{1}\left(x_{n-2}x_{n+1}+x_{n-1}x_{n+2}\right)
\label{eq:NP-HSection}
\end{eqnarray}
Following the same procedure as in Eqs.~(\ref{eq:activities}-\ref{eq:3RG})
yields the RG recursions
\begin{eqnarray}
\kappa' & = & \kappa\lambda\,\frac{\left(1+\mu\right)^{2}}{\left(1+\mu\kappa\right)^{2}},\nonumber \\
\lambda' & = & \frac{\left(\kappa+\mu\right)^{2}}{\left(1+\mu\kappa\right)^{2}},\label{eq:NP-RG}\\
C' & = & C^{2}\,\frac{\kappa\mu^{2}}{\left(1+\mu\right)^{2}\left(\kappa+\mu\right)\left(1+\mu\kappa\right)},\nonumber 
\end{eqnarray}
with the same initial conditions as for HN3 in Eq.~(\ref{eq:HN3activitiesIC}).

Once more inspecting these recursions for fixed points, $\kappa'=\kappa=\kappa^{*}$
and $\lambda'=\lambda=\lambda^{*}$, we find a strong-coupling line
of fixed points at 
\begin{equation}
\kappa^{*}=0,\quad\lambda^{*}=\mu^{2},\label{eq:NPRGfixedpointT0}
\end{equation}
which extends over the entire domain for the long-range bonds, $0\leq\lambda^{*}\leq1$
for $0\leq\mu\leq1$. We can similarly identify a high-temperature
fixed point at 
\begin{eqnarray}
\kappa^{*} & = & \lambda^{*}=1\label{eq:HNNPhighTfp}
\end{eqnarray}
that is valid for all $\mu$. Further analysis of Eqs.~(\ref{eq:NP-RG})
reveals yet another line of fixed points given by 
\begin{equation}
\kappa^{*}=\frac{1-\mu-\mu^{2}}{\mu^{2}},\quad\lambda^{*}=\frac{\left(1-\mu\right)^{2}}{\mu^{2}}.\label{eq:NPRGfpl}
\end{equation}
It is clear that this line of fixed points is physical merely in the
range $\frac{1}{2}\leq\mu\leq\frac{1}{\phi}$ beyond which one or
both of $\kappa^{*}$ and $\lambda^{*}$ leaves the unit interval.
We can express this line in closed form,
\begin{eqnarray}
\kappa^{*} & = & \lambda^{*}+\sqrt{\lambda^{*}}-1,\label{eq:NPkaplam_fpl}
\end{eqnarray}
by eliminating $\mu$. 

\begin{figure}
\includegraphics[clip,scale=0.33]{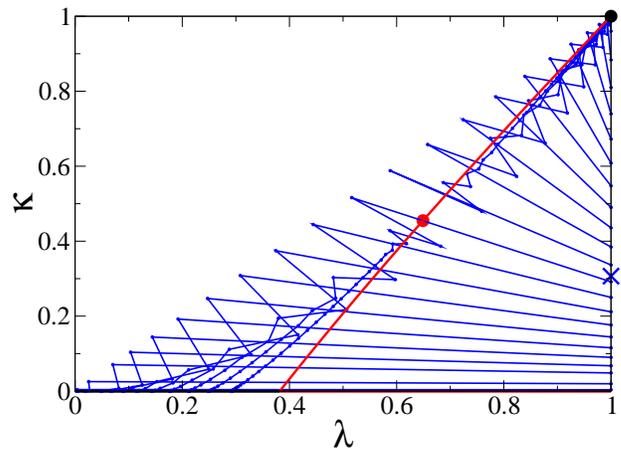}
\caption{(Color Online)\label{fig:IsingPhaseHNNP}Phase diagram for the Ising ferromagnet
on HNNP. Similar to HN3 in Fig.~\ref{fig:IsingPhaseHN3}, there is
again a high-temperature fixed point here (black dot). Though, more
akin to HN5 in Fig.~\ref{fig:IsingPhaseHN5}, a line of fixed points
extends along $\kappa^{*}=0$ and $0\leq\lambda^{*}\leq1$, which
for $\lambda^{*}<\phi^{-2}$ is stable (thick black line) and unstable
above (thick red line). Both regimes are connected by a (red) line
of unstable fixed points given by Eq.~(\ref{eq:NPkaplam_fpl}) with
an intercept at $\lambda=\phi^{-2}=0.381966\ldots$ For the homogeneous-bond
initial conditions in Eq.~(\ref{eq:HN3activitiesIC}), the ensuing
RG flow (blue lines) starts from $\lambda^{(0)}=1$, evolves toward
the diagonal $\kappa^{(1)}=\lambda^{(1)}$ first, and proceeds in
an oscillatory manner. For initial temperatures below the critical
point $\mu_{c}^{2}=0.30671\ldots$ (${\bf \times}$), the RG flow
eventually veers toward strong coupling, $\kappa^{*}=0$, and above,
to the high-temperature fixed point. The particular fixed point governing
the flow for these initial conditions is marked by a red dot.}
\end{figure}

We have plotted the phase diagram for HNNP in Fig.~\ref{fig:IsingPhaseHNNP}.
It also shows the RG flow for the initial conditions in Eqs.~(\ref{eq:HN3activitiesIC}),
which starts on the right-hand vertical axis, $\lambda^{(0)}=1$.
For these initial conditions, there appears to be an ordinary, finite-order
phase transition, with the RG-flow diverging from an unstable fixed
point toward a solidly frozen ordered regime at lower temperatures
and a plain disordered regime above. Yet, the transition is non-universal,
for instance, dependent on the initial ratio between $K_{0}$ and
$K_{1}$ bonds. There is no patchy order in this system, as there
are apparently not enough long-range bonds to enforce it, compared
to HN5. The actual critical temperature $T_{c}$ is the transcendental
solution of an infinite integration of the RG-recursions in Eq.~(\ref{eq:NP-RG})
from the initial conditions in Eq.~(\ref{eq:HN3activitiesIC}), which
can be obtained to any accuracy by a shooting procedure. With that,
we obtain 
\begin{eqnarray}
\mu_{c}=0.553814426157623\ldots,\label{eq:HNNPmuc}
\end{eqnarray}
or
\begin{eqnarray}
 &  & T_{c}=-\frac{2}{\ln\mu_{c}}=3.3845207164300986\ldots,\label{eq:HNNPTc}
 \end{eqnarray}
marked by a blue cross in Fig.~\ref{fig:IsingPhaseHNNP}. Inserted
into Eqs.~(\ref{eq:NPRGfpl}), this places the unstable fixed point
governing the critical behavior for these particular initial conditions
at $\left(\kappa^{*},\lambda^{*}\right)=\left(0.4547454105\ldots,0.64908641578\ldots\right)$,
labeled by a red dot. 

\subsubsection{Fixed-Point Stability\label{sub:Fixed-Point-Stability-HNNP}}

We explore the stability of these fixed points with a local analysis,
using again the Ansatz in Eq.~(\ref{eq:Ansatzdeleps}). First, we
consider the strong-coupling fixed point in Eq.~(\ref{eq:NPRGfixedpointT0}).
Expanding the RG-recursion for $\kappa'$ with the Ansatz yields
\begin{eqnarray}
\epsilon_{n+1} & = & \mu^{2}\left(1+\mu\right)^{2}\epsilon_{n},\label{eq:epsHNNP}
\end{eqnarray}
immediately proving that this fixed point is stable \emph{only} for
$\mu<\frac{1}{\phi}$, where the $\kappa=0$-line intercepts with
the other line of fixed points in Eqs.~(\ref{eq:NPRGfpl}) or (\ref{eq:NPkaplam_fpl}).
For values above that, $\kappa^{*}=0$ becomes unstable. Conversely,
expanding around the high-temperature fixed point in Eq.~(\ref{eq:HNNPhighTfp})
gives
\begin{eqnarray}
\epsilon_{n+1} & = & \frac{1-\mu}{1+\mu}\,\epsilon_{n}+\delta_{n},\nonumber \\
\delta_{n+1} & = & \frac{2\left(1-\mu\right)}{1+\mu}\,\epsilon_{n},
\label{eq:HNNPhighTeigensystem}
\end{eqnarray}
which possesses the eigenvalues
\begin{equation}
\alpha_{\pm} =\frac{1-\mu}{2\left(1+\mu\right)}\left[1\pm\sqrt{1+8\frac{1+\mu}{1-\mu}}\right].
\label{eq:HNNPhighTeigenvalues}
\end{equation}
The lower branch for all $0<\mu<1$ remains confined to $-1<\alpha_{-}<0$,
while $1<\alpha_{+}<2$ for $\mu<\frac{1}{2}$; only for $\mu>\frac{1}{2}$
is this fixed point stable. In the overlap, $\frac{1}{2}<\mu<\frac{1}{\phi}$,
of the stable regimes of the two previous fixed points, we find the
remaining line of fixed points in Eqs.~(\ref{eq:NPRGfpl}), which
prove to be unstable. To show that, we insert the Ansatz to obtain
the system
\begin{eqnarray}
\epsilon_{n+1} & = & \frac{\mu^{2}+2\mu-1}{1-\mu^{2}}\,\epsilon_{n}+\frac{1-\mu-\mu^{2}}{\left(1-\mu\right)^{2}}\,\delta_{n},\nonumber \\
\delta_{n+1} & = & \frac{2\mu}{1+\mu}\,\epsilon_{n}.\label{eq:HNNPeigensystem}
\end{eqnarray}
The eigenvalues $\alpha_{\pm}$ of this system are given by
\begin{eqnarray}
\alpha_{\pm} & = & \frac{\mu^{2}+2\mu-1\pm\sqrt{1+4\mu+2\mu^{2}-12\mu^{3}-7\mu^{4}}}{2\left(1-\mu^{2}\right)},\label{eq:HNNPeigenvalues}
\end{eqnarray}
for which $\alpha_{+}>1$ and $0>\alpha_{-}>-1$ on the physically
relevant interval $\frac{1}{2}<\mu<\frac{1}{\phi}$. Thus, any fixed
point on that line has an unstable direction and a stable but \emph{oscillatory}
direction, as is apparent for the flow in Fig.~\ref{fig:IsingPhaseHNNP}.

\subsubsection{Divergence of the Correlation Length at $T_{c}$\label{sub:Divergence-xi-HNNP}}

Although we have found that the phase diagram for HNNP consists of
lines of stable and unstable fixed points, only one specific unstable
fixed point controls the scaling behavior for any specific incarnation
of the system. Any such incarnation is selected by the initial choice
of couplings $\left(\kappa^{(0)},\lambda^{(0)}\right)$ as a function
of temperature $\mu$. The controlling fixed point in Eq.~(\ref{eq:NPRGfpl})
is a non-trivial consequence of these initial conditions. It is in
particular a function of the specific critical temperature $\mu_{c}$
for which it is reached. The induced scaling behavior for nearby temperatures
is thus non-universal. Yet, for a given $\mu_{c}$, the local analysis
for such an unstable fixed point proceeds identically to that for
Ising systems on a lattice \citep{Plischke94}. Here we have two scaling
fields, whose fixed-point analysis leads to two eigenvalues in Eq.
(\ref{eq:HNNPeigenvalues}) that yields one relevant and one irrelevant
variable. From the eigenvalue $\alpha_{+}\left(>1\right)$ for the
relevant variable, we obtain for the correlation-length exponent
\begin{equation}
\nu=\frac{1}{y_{+}}=\frac{\ln2}{\ln\alpha_{+}\left(\mu_{c}\right)}\approx10.8301
\ldots,\label{eq:nuHNNP}
\end{equation}
such that
\begin{equation}
\xi\sim\left|\mu-\mu_{c}\right|^{-\nu}.\label{eq:xi_nu}
\end{equation}
Thus, we find that the correlation length on approach to the fixed
point merely diverges with a power law here. In this particular case,
the exponent $\nu$ happens to be very large, such that it would be
difficult  to distinguish this divergence from the exponential
kind numerically.

\begin{figure}
\includegraphics[bb=0bp 0bp 300bp 209bp,clip,scale=0.8]{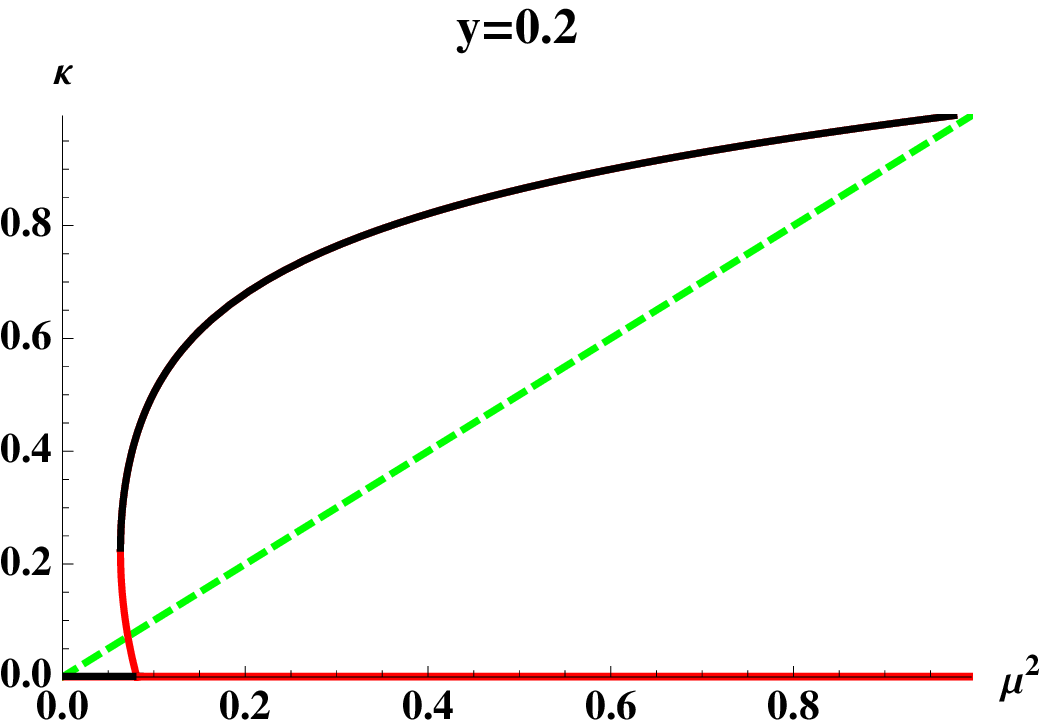}
\includegraphics[bb=0bp 0bp 300bp 209bp,clip,scale=0.8]{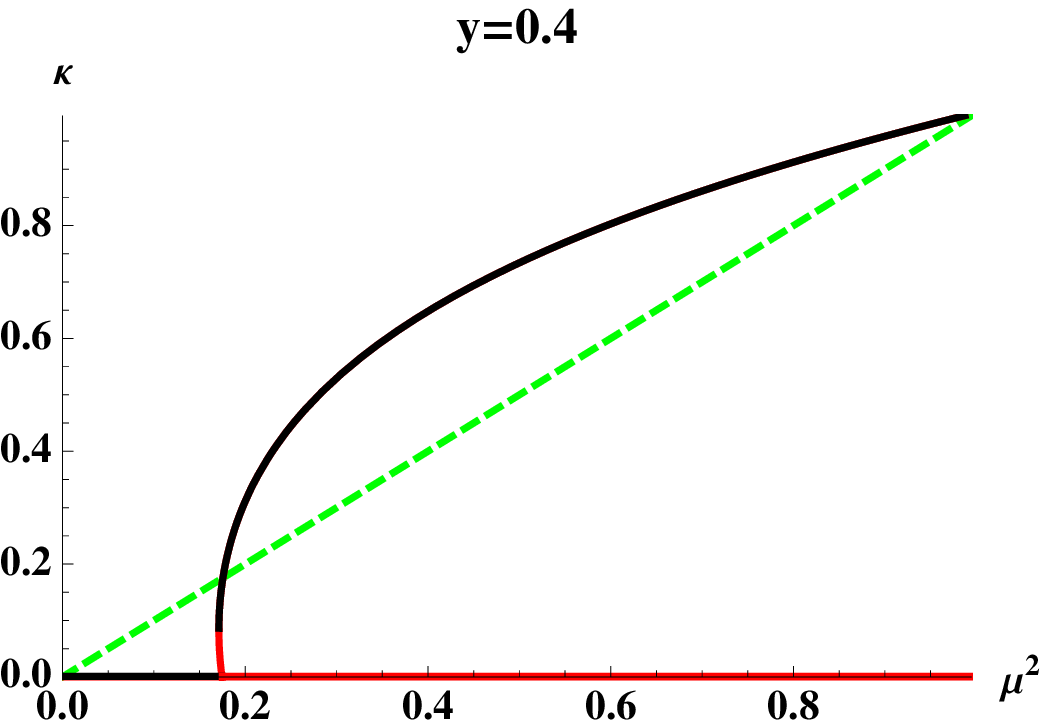}
\includegraphics[bb=0bp 0bp 300bp 209bp,clip,scale=0.8]{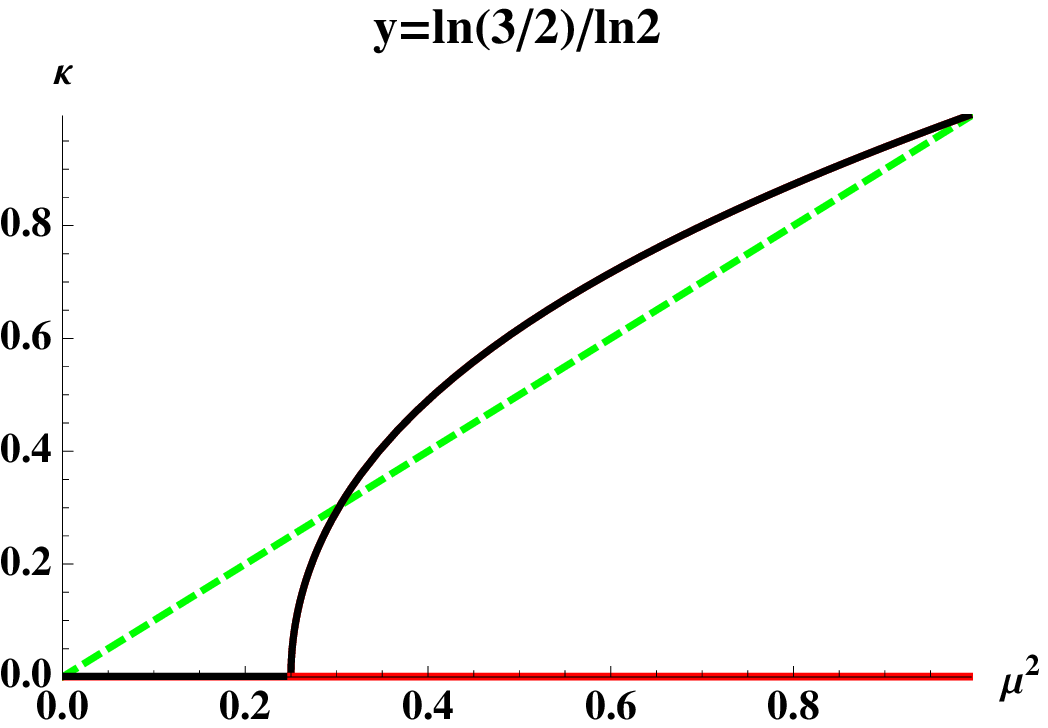}
\caption{(Color Online)\label{fig:HN5yPlots}Plot of $\kappa^{*}$ in Eq.~(\ref{eq:5yRGfpl})
as a function of $\mu^{2}$ for $y$ as defined in Eq.~(\ref{eq:L1bond})
for the interpolation between HN3 and HN5.
The generic cases are represented by $y=0.2$ (top), $y=0.4$ (middle),
and the special value of $y=\frac{\ln\left(3/2\right)}{\ln2}=0.58\ldots$
(bottom) at which the branch point (BP) {}``sunsets'' out of the
physical regime and the behavior becomes similar to that for $y=1$
shown in Fig.~\ref{fig:IsingPhaseHN5}. In all cases, the dashed
line indicates the initial conditions (IC) for the RG
flow in Eqs.~(\ref{eq:5RG}-\ref{eq:HN5activitiesIC}), $\kappa^{(0)}=\mu^{2}$.
For $\mu$ fixed, the RG flow \emph{must} proceed vertically, either
up or down, to the nearest stable line of fixed points. For low $y$
(top), the IC cross the unstable branch below BP which then can \emph{not}
be reached by the flow. Once the IC cross above BP (middle), the flow
\emph{must} pass BP, unless BP sunsets (bottom). }
\end{figure}

\section{Interpolation between Hanoi Networks\label{sec:HN5y}}

We have already observed in the construction of HN5 in
Sec.~\ref{sub:Ising-Ferromagnet-on-HN5} that it is easy to promote the
$L$-couplings, which inevitably emerge during the RG, to be associated
with an actual bond in the network.  Here, we will fully exploit this
fact to obtain a one-parameter family of problems with various regimes
of phase behaviors, based on the interpolating parameter $y$ in
Eq.~(\ref{eq:L1bond}).  This procedure has also been used in
Ref.~\citep{Boettcher11b} and a similar construction can already be
found in Ref.~\citep{Hinczewski07}. In particular, we discover
transitions between such regimes as a function of the parameter that
will allow us to clarify the connections between the diverse set of
behaviors that we have discovered in the previous section.

\subsection{Interpolation between HN3 and HN5\label{sub:Interpolation-HN3-HN5}}

In Sec.~\ref{sub:Ising-Ferromagnet-on-HN5}, we argued for the
introduction of small-world bonds with couplings $L_{i}$ and developed
the RG recursions in (\ref{eq:5RG}) assuming a relative strength of
these couplings to those germane to HN3 of the form in
Eq.~(\ref{eq:L1bond}).  Here, we will now consider the behavior that
results from varying the strength parameter $y$ between the two
extremes already explored, $y=0$ for HN3 in
Sec.~\ref{sub:Ising-Ferromagnet-on-HN3} and $y=1$ for HN5 in
Sec.~\ref{sub:Ising-Ferromagnet-on-HN5}.

Inspecting these recursions in Eq.~(\ref{eq:5RG}) for fixed points, we
already found the low-temperature fixed point line in
Eq.~(\ref{eq:5RGfixedpointT0}).  Aside from this
$\kappa^{*}=0$-solution, Eqs.~(\ref{eq:5RG}) further reveal a line of
fixed points given by
\begin{eqnarray}
\kappa^{*} & = & \frac{1}{2}\left[\mu^{y}\left(1+\mu\right)-2\mu\pm\sqrt{{\cal D}_{y}\left(\mu\right)}\right],\nonumber \\
\label{eq:5yRGfpl}\\
\lambda^{*} & = & \frac{\mu^{y}}{4}\left[2\left(1-\mu\right)+\mu^{y}\left(1+\mu\right)\pm\sqrt{{\cal D}_{y}\left(\mu\right)}\right],\nonumber 
\end{eqnarray}
abbreviating the discriminant 
\begin{equation}
{\cal  D}_{y}\left(\mu\right)=\left(1+\mu\right)\left[\mu^{2y}\left(1+\mu\right)-4\left(1-\mu^{y}\right)\left(1-\mu\right)\right].
\label{eq:Dy}
\end{equation}

For $y\to0$, this solution morphs into the high-temperature
fixed point of HN3, see Fig.~\ref{fig:IsingPhaseHN3}. But for any
finite $y$, these lines of fixed points are non-trivial functions
of $\mu$, as depicted for $\kappa^{*}\left(\mu^{2}\right)$ in Fig.
\ref{fig:HN5yPlots}. The dominant feature in these plots is the root-singularity
in $\kappa^{*}$ with a branch point separating the upper stable and
lower unstable line of fixed points. Essentially, three distinct generic
regimes can be discerned: (1) If the branch-point happens to lie below
the physical regime, we observe a phase transition without access
to any unstable point (see bottom of Fig.~\ref{fig:HN5yPlots}); a
critical point akin to that for HN5 at $y=1$  in Sec.~\ref{sub:Ising-Ferromagnet-on-HN5}
arises. If the branch point rises into the physical regime, here for
$y<y_{c}=\frac{\ln\left(3/2\right)}{\ln2}=0.584963\ldots$, then depending
on whether the initial conditions of the RG flow cross the critical
line below or above the branch point, we find (2) a transition seemingly
of finite-order on intercepting the unstable lower branch (see top
of Fig.~\ref{fig:HN5yPlots}) for which the RG flow never accesses
the branch point singularity. If, in turn, the initial conditions
cross above, (3) a BKT-like transition results because the RG flow
now \emph{must} pass the singularity (see middle of Fig.~\ref{fig:HN5yPlots}),
as we will show below.

\subsubsection{Fixed-Point Stability\label{sub:Fixed-Point-StabilityHN5y} }

As before, we determine the stability of the fixed points with a local
analysis using  Eq.~(\ref{eq:Ansatzdeleps}). Inserting the Ansatz for
the strong-coupling solution, Eq.~(\ref{eq:5RGfixedpointT0}),
yields
\begin{eqnarray}
\epsilon_{n+1} & = & \mu^{2y}\left(1+\mu\right)^{2}\epsilon_{n}\label{eq:epsHN5y}
\end{eqnarray}
and $\delta_{n}\propto\epsilon_{n}$. Thus, this line of fixed points
is stable only for $\mu$ smaller than the solution of $1=\mu^{y}\left(1+\mu\right)$;
for larger values the line is unstable. In each plot
of Fig.~\ref{fig:HN5yPlots}, that value of $\mu$ corresponds to
the location on the $\kappa=0$-line where the non-trivial fixed-point
line intercepts. 

Applying the Ansatz on the line of high-temperature fixed points in
Eqs.(\ref{eq:5yRGfpl}) obtains now 
\begin{eqnarray}
\epsilon_{n+1} & = & \frac{\left(1+\mu\right)\left(\mu^{y}-2\right)\pm\sqrt{{\cal D}_{y}\left(\mu\right)}}{2\left(1+\mu\right)\mu^{y}}\,\epsilon_{n}\nonumber \\
 & + & \frac{2\left(1-\mu\right)-\mu^{y}\left(1+\mu\right)\pm\sqrt{{\cal D}_{y}\left(\mu\right)}}{2\left(1-\mu\right)\mu^{y}}\,\delta_{n},\nonumber \\
\delta_{n+1} & = & \frac{\left(1+\mu\right)\mu^{y}\mp\sqrt{{\cal D}_{y}\left(\mu\right)}}{2\left(1+\mu\right)}\,\epsilon_{n}.\label{eq:HN5ymatrix}
\end{eqnarray}
It is tedious but straightforward to show that the two eigenvalues
$\alpha_{1,2}$ characterizing this system indicate an upper stable
branch for {}``+'' and a lower unstable branch for {}``-''. Yet,
the decisive question for the observed phase transition concerns the
intercept with the initial conditions of the RG flow, i.e. the relative
strengths of bare couplings in the network. If the intercept lies below
the branch-point singularity, the phase transition is similar to that
discussed in Sec.~\ref{sub:Ising-Ferromagnet-on-HNNP}, although here
the high-temperature phase flows not into a single point but rather
into a temperature-dependent line of couplings, similar to (but inverted
from) BKT. Apparently, such temperature-dependent couplings in themselves
do not imply an infinite-order transition. 

In turn, once those initial conditions pass \emph{above} the branch
point, such as in the middle panel of Fig.~\ref{fig:HN5yPlots}, there
is no stable fixed point above that and the flow has to move toward
lower values of $\kappa$ during the RG flow. Whether there is any
intercept with the upper branch makes little difference to the generic
case here, as the critical behavior is controlled now by the flow in the
immediate neighborhood of the root-singularity where both branches
pinch off. Only in this case do we obtain an infinite-order transition
of the BKT type.

\subsubsection{Divergence of the Correlation Length at $T_{c}$\label{sub:Divergence-HN5y}}

We have already considered the exponential divergence characteristic
of HN5 for $y>y_{c}$ for the special case $y=1$ in Sec.~\ref{sub:Exponential-Divergence-of}.
Similarly, we have found in Sec.~\ref{sub:Divergence-xi-HNNP} that
HNNP at $y=1$ has a critical point very much like that found for
HN5 at small $y>0$, aside from a difference in the high-temperature
coupling. For both cases, shown in the bottom and top panel of Fig.~\ref{fig:HN5yPlots}, respectively, further analysis would not alter the
qualitative behavior in the divergence of the correlation length $\xi$. 

Therefore, we restrict ourselves here to the novel case where the
RG flow must pass near the branch-point singularity for a given choice
of initial conditions displayed in the middle panel of Fig.~\ref{fig:HN5yPlots}
for HN5 with $y$ just below $y_{c}$. As has been argued for a similar
case in a scale-free version of the hierarchical lattice \citep{Hinczewski06},
for temperatures $\mu$ approaching the critical temperature $\mu_{c}$
ever closer from below, the flow spends an ever larger number of iterations
near the singularity. This characteristic number of RG iterations
$n^{*}\left(\mu\right)$ diverges for $\mu\to\mu_{c}^{-}$ and can
be interpreted as a correlation measure, just as in Sec.~\ref{sub:Exponential-Divergence-of}.
But unlike the $n^{*}\sim1/\left|\mu_{c}-\mu\right|$ scaling there,
we find here that $n^{*}\sim1/\sqrt{\left|\mu_{c}-\mu\right|}$, leading
the the BKT-like transition also found in many other networks.

To derive $n^{*}$, we follow Ref.~\citep{Hinczewski06} in ignoring
the transient number of RG steps that would take us from the initial
conditions to the singularity. Instead, we imagine that our flow would
start right at the singularity but with an infinitesimally small shift
down in temperature, 
\begin{equation}
\tau=\mu_{c}-\mu,
\label{eq:def_tau}
\end{equation}
so as to dislodge the flow from what is a fixed point for $\mu=\mu_{c}$.
The system then decorrelates, i.e. the RG flow escapes the singularity
by a finite amount, for sizes $N=2^{n}$, i.e. after $n>n^{*}$ iterations
of the RG. 

As the transition is obtained right at the branch point, the critical
temperature $\mu_{c}$ derives from
\begin{equation}
{\cal D}_{y}\left(\mu_{c}\right)=0
\label{eq:Def_muc}
\end{equation}
in Eqs.~(\ref{eq:5yRGfpl}-\ref{eq:Dy}). We define the critical-point couplings
\begin{equation}
\left[\begin{array}{c}
\kappa_{c}\\
\lambda_{c}\end{array}\right]=\left[\begin{array}{c}
\kappa^{*}\left(\mu_{c}\right)\\
\lambda^{*}\left(\mu_{c}\right)\end{array}\right]=\left[\begin{array}{c}
-\mu_{c}+\frac{\mu_{c}^{y}}{2}\left(1+\mu_{c}\right)\\
\frac{\mu_{c}^{y}}{2}\left(1-\mu_{c}\right)+\frac{\mu_{c}^{2y}}{4}\left(1+\mu_{c}\right)\end{array}\right]
\label{eq:kc_lc}
\end{equation}
from Eq.~(\ref{eq:5yRGfpl}). Such a phase transition only exists
for both, $\kappa_{c}$ and $\lambda_{c}$, in the physical regime
and, hence, it disappears when $\kappa_{c}=0$ is reached. The simultaneous solution
of $\kappa_{c}=0$ and Eq.~(\ref{eq:Def_muc}) yield 
\begin{equation}
y_{c}=\frac{\ln\frac{3}{2}}{\ln2}\qquad{\rm  at\qquad}\mu_{c}=\frac{1}{2},
\label{eq:yc_HN5}
\end{equation}
see Fig.~\ref{fig:HN5yPlots} (bottom), with a transition similar
to that at $y=1$ for all $y>y_{c}$. 

For $y<y_{c}$, we  expand Eqs.~(\ref{eq:5RG}) around the branch point 
to \emph{2nd} order in $\epsilon_{n}$, $\delta_{n}$ with \[
\left[\begin{array}{c}
\kappa_{n}\\
\lambda_{n}\end{array}\right]=\left[\begin{array}{c}
\kappa_{c}+\epsilon_{n}\\
\lambda_{c}+\delta_{n}\end{array}\right]\]
and $\epsilon_{0}=\delta_{0}=0$. Note that neither
$\kappa_{c}$ nor $\lambda_{c}$ depend on $\mu$; it is precisely
the \emph{explicit} appearance of $\mu$ in Eqs.~(\ref{eq:5RG}) that
allows a further \emph{1st}-order expansion in $\tau$ according to
Eq.~(\ref{eq:def_tau}). For general $y$, the solutions of Eq.~(\ref{eq:Def_muc})
are too messy to display the results, but the nature of the recursions
is as follows:
\begin{eqnarray}
\epsilon_{n+1} & \sim & A\tau+B\epsilon_{n}+C\delta_{n}+D\epsilon_{n}^{2}+E\epsilon_{n}\delta_{n},\nonumber \\
\delta_{n+1} & \sim & F\tau+G\epsilon_{n}+H\epsilon_{n}^{2},
\label{eq:ed_rec}
\end{eqnarray}
dropping terms cubic in $\epsilon,\delta$ and any smaller terms in
$\tau$. All coefficients are algebraic functions of $\mu_{c}$ and $y$.
The proper scaling field $u_{n}$ for the impending analysis is identified
by multiplying the $\delta$-equation by $C$ and combining it with
that of $\epsilon$:
\begin{eqnarray}
u_{n+1} & = & \epsilon_{n+1}+C\delta_{n+1},\label{eq:u_rec}\\
 & \sim &
\left(A+CF\right)\tau+\left(B+CG\right)\epsilon_{n}+C\delta_{n}\nonumber\\
&&\quad+\left(D+CH\right)\epsilon_{n}^{2}+E\epsilon_{n}\delta_{n}.\nonumber 
\end{eqnarray}
The subtlety of this procedure reveals itself now in the fact that
$B+CG\equiv1$ for any $\mu_{c}$ and $y$, and we can write with some
new (positive) coefficients\[
u_{n+1}-u_{n}\sim u'\left(n\right)\sim-P\tau-Qu_{n}^{2},\]
using a continuum approach justified for large $n$. In the limit
$\tau\to0$, this equation describes a singular boundary layer problem
easily analyzed with the techniques outlined in Ref.~\citep{BO}.
Instead, we note that the differential equation can be solved exactly
to give (with $u_{0}=0$):
\begin{equation}
u_{n}\sim-\sqrt{\frac{P}{Q}\tau}\tan\left(\sqrt{PQ\tau}n\right).\label{eq:un_sol}
\end{equation}
Thus, the solution exits the boundary layer near the branch point
(moving down toward the strong-coupling fixed point) and becomes
finite for $\sqrt{PQ\tau}n^{*}\to\pi/2$, and we identify the scaling
\begin{equation}
n^{*}\sim\frac{1}{\sqrt{\tau}}=\frac{1}{\sqrt{\mu_{c}-\mu}},\label{eq:nstar-HN5y}
\end{equation}
which by Eq.~(\ref{eq:def_xi}) leads to the divergence in the correlation
length characteristic of BKT,
\begin{equation}
\xi\left(\mu\right)\sim e^{\frac{const}{\sqrt{\mu_{c}-\mu}}},\qquad\mu\to\mu_{c}^{-}.\label{eq:xi_BKT}
\end{equation}
Clearly, the physical origin of this singularity is not even remotely
related to an actual BKT transition. In fact, instead of its rarity,
confined to very particular lattice models, we may find it to be one
of a few generic types of transition in networks.

\begin{figure}[b!]
\includegraphics[clip,scale=0.8]{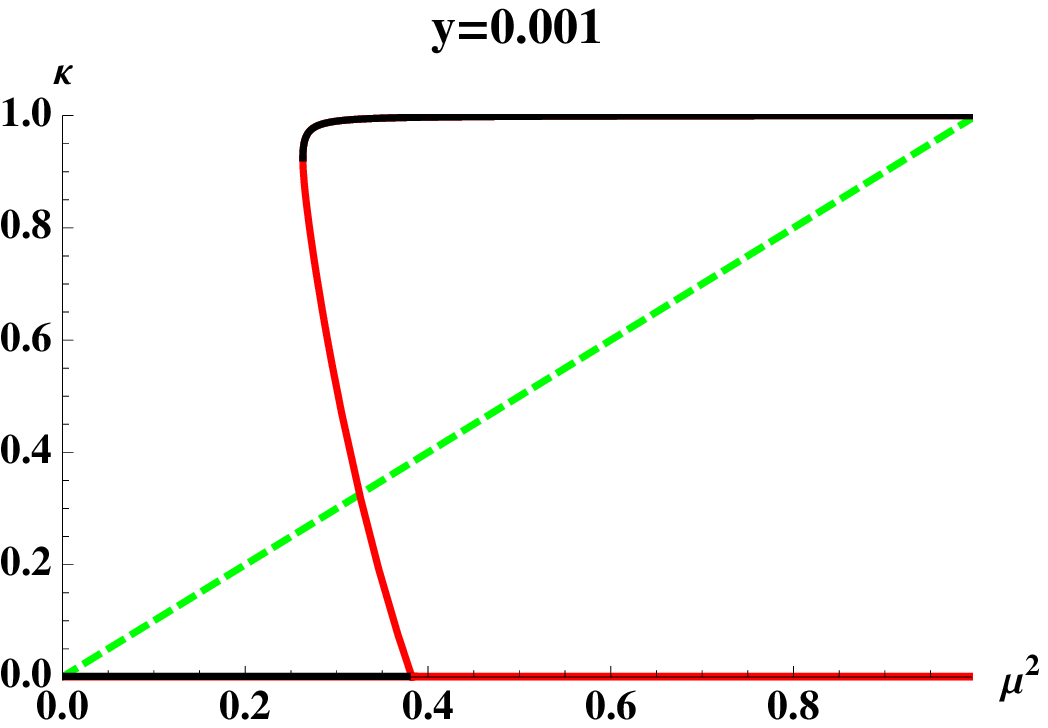}
\includegraphics[clip,scale=0.8]{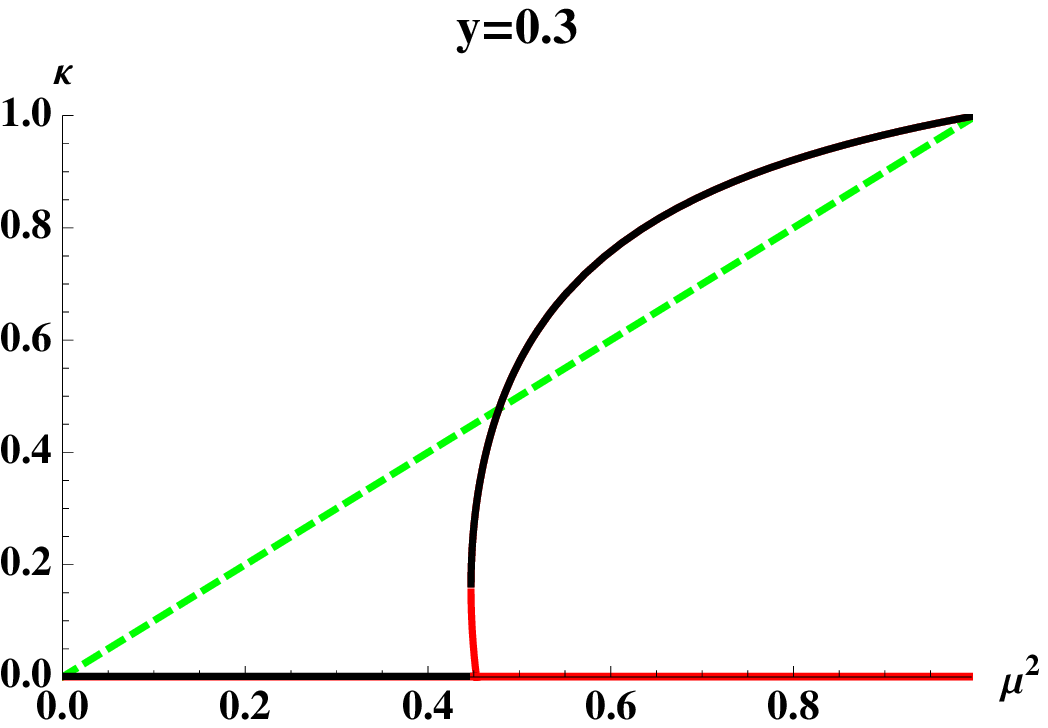}
\includegraphics[clip,scale=0.8]{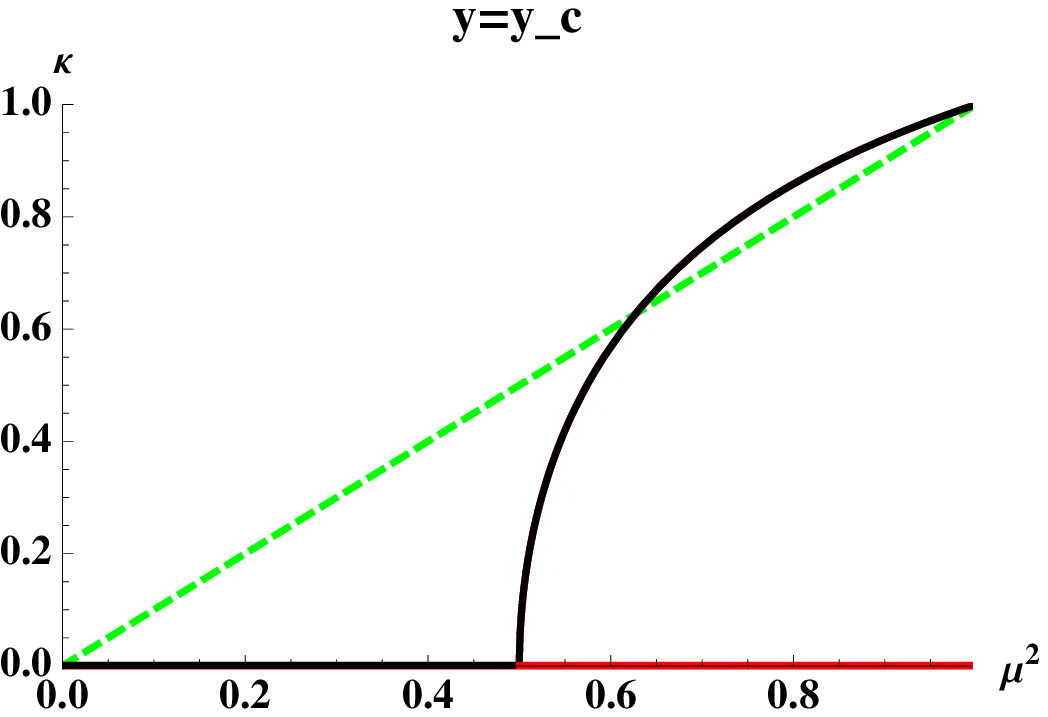}
\caption{(Color Online)\label{fig:kstarHNNP}Plot of $\kappa^{*}$ in Eq.~(\ref{eq:HN6highTfp})
as a function of $\mu^{2}$ for $y$ as defined in Eq.~(\ref{eq:L1bond})
for the interpolation between HNNP and HN6,
similar to Fig.~\ref{fig:HN5yPlots}. The generic cases are represented
by $y\approx0$ (top), $y=0.3$ (middle), and the special value of
$y=2\left[\frac{\ln\left(1+\sqrt{2}\right)}{\ln2}-1\right]=0.54\ldots$
(bottom) at which the branch point {}``sunsets'' out of the physical
regime, see discussion in Fig.~\ref{fig:HN5yPlots}.}
\end{figure}

\subsection{Interpolation between HNNP and HN6\label{sub:Analysis-of-HN6:} }

Similar to the step from HN3 to HN5, we can add the emerging bonds
$L_{1}$ again as an original feature of the network, which we parameterize
as in Eq.~(\ref{eq:L1bond}). As a result, we obtain a network that
has an average degree of 6 with an exponential distribution of bonds
which we will call HN6, accordingly. While difficult to display, HN6
consists of the addition of the green-shaded bonds from HN5 in Fig.
\ref{fig:5hanoi} to HNNP in Fig.~\ref{fig:hanoi_nonplanar}. 

As in Eq.~(\ref{eq:L1bond}), we add the extra $L$-bonds with an
initial coupling strength that is a $y$-fraction of the original
HNNP bonds; for $y\to0$ we approach the pure HNNP from before, and
for $y=1$ we attain HN6 with all equal couplings. In this calculation,
we get similar to Eqs.~(\ref{eq:5RG}) and (\ref{eq:NP-RG}):
\begin{eqnarray}
\kappa' & = & \kappa\lambda\,\frac{\left(1+\mu\right)^{2}}{\left(1+\mu\kappa\right)^{2}},\nonumber \\
\lambda' & = & \mu^{2y}\frac{\left(\kappa+\mu\right)^{2}}{\left(1+\mu\kappa\right)^{2}},\label{eq:HN6RG}\\
C' & = & C^{2}\,\frac{\kappa\mu^{2}}{\left(1+\mu\right)^{2}\left(\kappa+\mu\right)\left(1+\mu\kappa\right)},\nonumber 
\end{eqnarray}
which is a simple extension of Eqs.~(\ref{eq:NP-RG}), to which these
RG recursions reduce for $y=0$. Similar to the interpolation between
HN3 and HN5, this interpolation between HNNP and HN6 also possesses
interesting phase behavior.
In particular, we also find a non-trivial value of $y_{c}$ separating
a HNNP-like phase diagram for low $y$ from one similar to HN5 for
larger $y$.

First, we find from Eqs.~(\ref{eq:HN6RG}) the strong-coupling fixed-point
line 
\begin{eqnarray}
\kappa_{0}^{*}=0, & \qquad & \lambda_{0}^{*}=\mu^{2+2y},\label{eq:HN6strongfp}
\end{eqnarray}
which is stable for all $\mu$ below the solution of
$1 =  \mu^{1+y}\left(1+\mu\right)$.
The remaining fixed points in the physical regime satisfy
\begin{eqnarray}
\kappa^{*} & = & -\frac{1}{\mu}+\frac{1+\mu}{2\mu^{2-y}}\left[1\pm\sqrt{\Delta_{y}\left(\mu\right)}\right],\label{eq:HN6highTfp}\\
\lambda^{*} & = & -\frac{1-\mu}{\mu^{1-y}}+\frac{1}{2\mu^{2-2y}}\left[1\pm\sqrt{\Delta_{y}\left(\mu\right)}\right],
\end{eqnarray}
where we abbreviate the discriminant
\begin{eqnarray}
\Delta_{y}\left(\mu\right) & = & 1-4\mu^{1-y}\left(1-\mu\right).\label{eq:DEFpsi}
\end{eqnarray}
Note that for $y\to0,$ the upper branches condense into the high-temperature
fixed point of HNNP, $\kappa^{*}=\lambda^{*}\to1$, and the lower
branch becomes the unstable fixed-point line of HNNP. It is similarly
easy to show that, indeed, for all $0\leq y\leq1$, {}``$+$'' corresponds
to stable fixed points and {}``$-$'' to unstable ones. 

At this point, it is obvious that the behavior for this system
parallels that for HN5 above. In Fig.~\ref{fig:kstarHNNP}, we display
for representative values of $y$ the dependence of $\kappa^{*}$ in
Eq.~(\ref{eq:HN6highTfp}) as a function of temperature $\mu$, which
in each panel is equivalent to those in Fig.~\ref{fig:HN5yPlots}: For
the smallest values of $y$ (here, including also $y=0$), the initial
conditions in Eq.~(\ref{eq:HN3activitiesIC}) result in an RG flow that
avoids the branch-point singularity and instead cross a line of
unstable fixed points, one of which dominates the critical behavior
that leads to a non-universal power-law divergence in the correlation
length $\xi$ as in Eq.~(\ref{eq:xi_nu}) for $y=0$.  For stronger
long-range couplings in the network, beyond some value of $y$ for
those initial conditions, that intercept moves above the branch point,
forcing the flow through it at the critical point, which now develops
a BKT-like essential singularity in $\xi$. Increasing such coupling
strength further moves the branch point below the physical regime, out
of reach from any initial condition. Here, this requires both,
$\Delta_{y}\left(\mu\right)$ in Eq.~(\ref{eq:DEFpsi}) and $\kappa^{*}$
in Eq.~(\ref{eq:HN6highTfp}), to vanish, which occurs for
\begin{eqnarray}
\mu_{c} & = & \frac{1}{\sqrt{2}},\label{eq:yc}\\
y_{c} & = & -\frac{\ln\left[\mu_{c}\left(1+\mu_{c}\right)\right]}{\ln\mu_{c}}=0.5431066\ldots.\nonumber 
\end{eqnarray}
Hence, for all $y>y_{c}$, there is no more high-temperature line
of unstable fixed points and HN6 behaves identical to HN5 in Sec.
\ref{sub:Ising-Ferromagnet-on-HN5}. 

The analysis of the critical behavior for the regimes of $y_{c}<y\leq1$
and for $y$ just below $y_{c}$ proceeds identically to Sec.~\ref{sub:Ising-Ferromagnet-on-HN5}
and Sec.~\ref{sub:Interpolation-HN3-HN5}, respectively. For the former,
we just mention that the expansion along the upper stable branch obtains
\begin{eqnarray}
\epsilon_{n+1} & = & \frac{\mu^{2}-\sqrt{\Delta_{y}\left(\mu\right)}}{1-\mu^{2}}\,\epsilon_{n}\nonumber \\
 &  & \quad+\left[\frac{\mu^{-y}}{1-\mu}-\frac{\mu\left(1-\sqrt{\Delta_{y}\left(\mu\right)}\right)}{2\left(1-\mu\right)^{2}}\right]\delta_{n},\nonumber \\
\delta_{n+1} & = & \frac{\mu^{y}}{1+\mu}\left[1-\sqrt{\Delta_{y}\left(\mu\right)}\right]\epsilon_{n}.\label{eq:linearHNNPy}
\end{eqnarray}
For example, at $y=1$, the behavior of the eigenvalues $\alpha_{\pm}$
of this system qualitatively resemble that in Fig.~\ref{fig:IsingHN5ev}:
The larger eigenvalue destabilizes (i.e. exceeds unity) when the temperature
is lowered to 
\begin{equation}
\mu_{c}=\frac{1}{3}\left[\sqrt[3]{\frac{25-3\sqrt{69}}{2}}+\sqrt[3]{\frac{25+3\sqrt{69}}{2}}-1\right]=0.754878\ldots\label{eq:mcHN6}
\end{equation}
or $kT_{c}/J=7.11239\ldots$, below which both eigenvalues disappear
separately into the complex plane at $\mu=\frac{3}{4}$. Near $\mu_{c}$,
we find for the dominant eigenvalue $\alpha_{+}\sim1-a\left|\mu-\mu_{c}\right|$,
with a constant $a=\frac{2}{3}\left[1+\sqrt[-3]{\left(623-75\sqrt{69}\right)/2}+\sqrt[3]{\left(623-75\sqrt{69}\right)/2}\right]=6.43855\ldots$,
leading to the same conclusion as in Sec.~\ref{sub:Exponential-Divergence-of}
with a correlation length given by Eq.~(\ref{eq:xi_exp-divergence}).

In turn, for values of $y$ just below $y_{c}$, we locate the branch
point via $\Delta_{y}\left(\mu_{c}\right)=0$ to obtain the critical
temperature $\mu_{c}(y)$ and find 
\begin{equation}
\left[\begin{array}{c}
\kappa_{c}\\
\lambda_{c}\end{array}\right]=\left[\begin{array}{c}
\kappa^{*}\left(\mu_{c}\right)\\
\lambda^{*}\left(\mu_{c}\right)\end{array}\right]=\left[\begin{array}{c}
\frac{1-2\mu_{c}^{2}}{\mu_{c}}\\
\frac{1-2\mu_{c}^{2}}{4\mu_{c}\left(1-\mu_{c}\right)^{2}}\end{array}\right],\label{eq:kappacHNNPy}
\end{equation}
where we have eliminate all explicit reference to $y$. Following
the steps subsequent to Eq.~(\ref{eq:kc_lc}), we find in the same
manner a BKT-like divergence of $\xi$ as in Eq.~(\ref{eq:xi_BKT}).

\section{Conclusion\label{sec:Conclusion}}

We have analyzed the fixed-point structure of an Ising ferromagnet on
a set of Hanoi networks with an exact real-space renormalization
group. Using interpolating families of such networks, with the
relative coupling strength $y$ between backbone and small-world bonds
as the interpolation parameter, we reveal a number of regimes with
distinct critical behaviors. While in each such regime the critical
transition has non-universal features, the characteristics of the
transition in each one has generic, robust features. For increasing
strength, we observe that the divergence in the correlation length
changes from power-law $x^{-\nu}$, to a BKT-like essential singularity
$e^{1/\sqrt{x}}$, then to a plain essential singularity $e^{1/x}$, on
approach to the critical point $x\sim\left|\mu_{c}-\mu\right|\to0$. We
trace the changes from one regime to the next in terms of the analytic
structure of the RG flow. Finding an enumerable range of such
characteristics hints at a possible classification of critical
behavior of statistical models in networks generally, a task we will
explore in more detail elsewhere \citep{Boettcher09b}. For example,
critical properties of the kind found here have also been observed in
community formation \citep{Hinczewski07} and in percolation
\citep{Boettcher09c,Berker09,Nogawa09,Hasegawa10a,Hasegawa10b}.  The
existence of entire regimes that exhibit essential singularities in
the divergence of the correlations, as we have found here, might
explain the surprising prevalence of typically quite rare BKT-like
transitions in otherwise unrelated network models
\citep{Dorogovtsev08}.

There are a number of possible extensions of this work, which we are
currently exploring. Here, we have merely focused on the fixed-point
structure and exclusively considered the divergence of the correlation
length. We are currently developing the RG scheme to investigate more
complicated observables to extract insights into the physical origin
of these peculiar critical phenomena. Merely the addition of an
external field as a generator for magnetic properties already makes
the analysis much more complicated by breaking the $Z_2$-symmetry,
thus allowing for yet another renormalizable bond linking three spins simultaneously
\citep{Boettcher11c}.  As well, the study of other models spin in this
context seems desirable. The absence of a bi-partite structure
(i.e.~the existence of odd-length loops) in the Hanoi networks
promises interesting effects due to frustration, which can be
calculated with similar rigor for an anti-ferromagnetic system (and, at
least, numerically exact for a spin glass).  The results of those
investigations are forthcoming.

\bibliography{/Users/stb/Boettcher}

\end{document}